\documentclass{aastex631}

\usepackage{natbib}
\usepackage{placeins}

\newcommand\cotwo{{CO$_2$}}
\newcommand\water{{H$_2$O}}

\newcommand\ajump{{$a$-jump}}

\newcommand\htwooproduction{{1.2$\times10^{24}$}}
\newcommand\coproduction{{5.2$\times10^{24}$}}
\newcommand\cotwoproduction{{(6.99$\pm$0.07)$\times 10^{24}$}}

\newcommand\deff{{$D_{eff.} =$ 5.9~$\mu$m}}
\newcommand\fice{{$f_{ice} = 33\%$}}

\received{June 11, 2025}
\accepted{May 1, 2026}

\submitjournal{PSJ}

\begin{document}

\title{JWST and Gemini Observations of the Active Centaur 450P/LONEOS: Nucleus and Coma Characterizations}

\correspondingauthor{C. A. Schambeau}
\email{charles.schambeau@ucf.edu}

\author[0000-0003-1800-8521]{Charles A. Schambeau}
\affiliation{Florida Space Institute, University of Central Florida, Orlando, FL, USA}
\affiliation{Department of Physics, University of Central Florida, Orlando, FL, USA}

\author[0000-0002-6702-7676]{Michael S. P. Kelley}
\affiliation{Department of Astronomy, University of Maryland, College Park, MD, USA}

\author[0000-0003-4659-8653]{Maria Womack}
\affiliation{Department of Physics, University of Central Florida, Orlando, FL, USA}

\author[0000-0002-7696-0302]{Eva Lilly}
\affiliation{Planetary Science Institute, Tucson, AZ, USA}

\author[0000-0003-1008-7499]{Theodore Kareta}
\affiliation{Department of Astrophysics and Planetary Science, Villanova University, Villanova, PA, USA}
\affiliation{Lowell Observatory, Flagstaff, AZ, USA}

\author[0000-0003-0194-5615]{Sara Faggi}
\affiliation{Department of Physics, American University, Washington D.C., USA}
\affiliation{NASA Goddard Space Flight Center, Greenbelt, MD, USA}

\author[0000-0002-2014-8227]{Olga Harrington Pinto}
\affiliation{Department of Physics, Auburn University, Auburn, AL, USA}

\author[0000-0001-7895-8209]{Marco Micheli}
\affiliation{ESA NEO Coordination Centre, Largo Galileo Galilei, 1, I-00044 Frascati (RM), Italy}

\author[0000-0002-8130-0974]{Dominique Bockel\'ee-Morvan}
\affiliation{LIRA, Observatoire de Paris, Universit\'e PSL, Sorbonne Universit\'e, Universit\'e Paris Cit\'e, Meudon, France}

\author[0000-0003-1156-9721]{Yanga R. Fern\'andez}
\affiliation{Department of Physics, University of Central Florida, Orlando, FL, USA}
\affiliation{Florida Space Institute, University of Central Florida, Orlando, FL, USA}

\author[0000-0002-0622-2400]{Adam McKay}
\affiliation{Department of Physics and Astronomy, Appalachian State University, Boone, NC, USA}

\author[0000-0002-2770-7896]{Noemi Pinilla-Alonso}
\affiliation{Department of Physics, University of Central Florida, Orlando, FL, USA}

\author[0000-0002-9214-337X]{Javier Licandro}
\affiliation{Instituto de Astrofísica de Canarias (IAC), 38205 La Laguna, Tenerife, Spain}
\affiliation{Departamento de Astrofísica, Universidad de La Laguna, 38206 La Laguna, Tenerife, Spain}

\author[0000-0002-5780-7062]{Aren Beck}
\affiliation{Department of Physics, University of Central Florida, Orlando, FL, USA}
\affiliation{Florida Space Institute, University of Central Florida, Orlando, FL, USA}

\author[0000-0002-2662-5776]{Geronimo L. Villanueva}
\affiliation{NASA Goddard Space Flight Center, Greenbelt, MD, USA}

\author[0000-0001-9542-0953]{James Bauer}
\affiliation{Department of Astronomy, University of Maryland, College Park, MD, USA}

\author[0000-0002-4230-6759]{Lori Feaga}
\affiliation{Department of Astronomy, University of Maryland, College Park, MD, USA}

\author{Michael A. DiSanti}
\affiliation{Solar System Exploration Division, Planetary Science Laboratory Code 693, NASA}
\affiliation{NASA Goddard Space Flight Center, Greenbelt, MD, USA}

\author{Kacper Wierzchos}
\affiliation{Lunar and Planetary Laboratory, University of Arizona, Tucson, AZ, USA}

\begin{abstract}

Between 2019 and 2024, we used the Gemini-N and JWST observatories to conduct a detailed case study of the active Centaur 450P/LONEOS, whose orbit was significantly altered by a close Saturn encounter in 1992. 
Gemini-N GMOS optical images likely captured the first views of 450P’s inactive nucleus, indicating a relatively small radius of $ R_N$~=~1.8$\pm$0.5~km and a surface color of $g'-i'$~=~1.15$\pm$0.09. 
This places 450P on the red end of the neutral/gray Centaur population and may indicate comparatively limited solar-driven surface processing relative to other known active Centaurs. 
A coma developed as 450P changed its heliocentric distance, $R_H$, from 7.83~au to 7.24~au, with an estimated low dust production rate of $\sim$~4-8~kg~s$^{-1}$.  
JWST NIRSpec IFU Prism-mode spectra revealed an elongated dust morphology and a symmetric \cotwo{} gas distribution in the coma but no H$_2$O or CO emission features, with production rates of  $Q_{\mathrm{CO_2}}$ = \cotwoproduction{} molec. s$^{-1}$, $Q_{\mathrm{H_2O}} \le$ \htwooproduction{} molec. s$^{-1}$, and $Q_{\mathrm{CO}} \le$ \coproduction{} molec. s$^{-1}$. 
Absorption features at 2.0 and 3.0~$\mu$m indicate the presence of water ice, and a subtle 3.1 $\mu$m feature, which is consistent with crystalline water ice in larger grains. 
A Hapke-style model dominated by large (\deff{}) dust grains with a volumetric ice fraction of \fice{} fits the spectrum. 
A thermal model incorporating 450P’s orbital history since $\sim$~1500~CE aligns with the observed onset of activity driven by \cotwo{} outgassing from amorphous water ice crystallization between 140–160~K.

\end{abstract}


\section{Introduction}
\label{sec:intro}

The solar system's Trans-Neptunian Objects (TNOs) and their dynamically linked descendants, the Centaurs\footnote{We adopt the definition for Centaurs in \cite{jewitt_2009}: an object's perihelion distance and semi-major axis are between the semi-major axes of Jupiter and Neptune and that are not in a 1:1 mean-motion resonance with a planet.} and Jupiter-family comets (JFCs), contain some of the most primitive materials remaining from the Solar System's formation. 
Their compositions can constrain physicochemical models of the early stages of planetesimal accretion, \citep{weidenschilling_1997_icarus, bergin-2024-comets-III, Simon-2024-comets-III, johansen2025}, including identifying the locations of snowlines that existed before dust was cleared from the midplane of the protoplanetary disk \citep{zhang-2024RvMG-snowlines}. 
However, establishing direct links between the present-day volatile inventories of these small icy bodies, after $\sim$ 4.5 billion years of material evolution and gravitationally induced orbital changes, remains challenging \citep{fraser-2024-comets-III} largely because the volatile abundances are measured for only a few Centaurs due to their relative faintness.
With the advent of the James Webb Space Telescope (JWST), detailed coma studies of Centaurs are now possible and these datasets provide strong observational constraints for thermophysical models of nucleus interiors.

Approximately 15\% of Centaurs display periods of activity \citep{jewitt_2009, fernandez2025}. 
This includes discrete outbursts (e.g., 29P/Schwassmann-Wachmann 1, 174P/Echeclus, C/2023 RS$_{61}$ \citep{lilly-2025RNAAS-RS61,kareta2025}), persistent heliocentric-distance-dependent behavior (e.g., 423P/Lemmon, 450P/LONEOS, 467P/LINEAR-Grauer), or an increase or reactivation in activity near aphelion counter to expectations based on diminishing solar heating (e.g., 95P/Chiron \citep{dobson-2024PSJ.....5..165D}).
It is unclear whether these different behaviors reflect fundamental differences in Centaur interiors arising from distinct TNO source populations or result from nuclei with originally similar compositions that experienced different amounts of thermal processing after entering Centaur orbits.
Notably, the nucleus thermal environment for all Centaurs is too cold for the vigorous sublimation of water ice, which is the dominate source of activity for comets within $\sim$ 3 au of the Sun \citep{womack_2017, bauer2025}.

Interestingly, all known active Centaurs have experienced a significant decrease in their semi-major axis ($a$) within the last $\sim$ 200 years, caused by a close encounter with a giant planet, defined as an ``$a$-jump" \citep{lilly-2024ApJ}. 
This inward migration increasingly heats the nucleus, potentially leading to phase transitions, a coma, and devolatilization of the surface and near-surface layers.
Centaurs that have recently undergone an $a$-jump are high-priority observational targets as they may have significant volatile inventories activated for the first time.
Models predict that Centaurs experience minimal surface erosion and that, unless another inward $a$-jump occurs, near-surface volatiles are gradually depleted, causing activity to cease or fall below current detection limits \citep{davidsson-2021thermal-model}.

\begin{figure}[ht]
\centering
\includegraphics[width=0.85\textwidth]{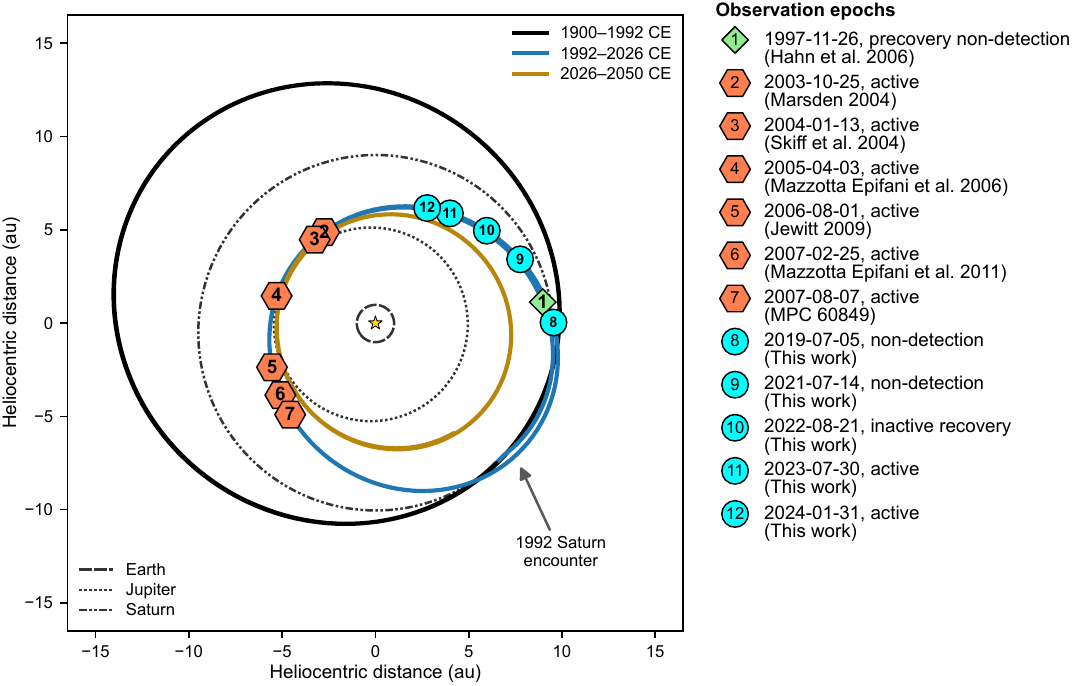}
\caption{450P's orbital history is visualized with this diagram. 
The Centaur had a close approach to Saturn in 1992 that changed its long-term trans-Saturnian  orbit (black curve) into one with a perihelion closer to Jupiter (blue). 
450P is predicted to have a close encounter with Jupiter in July 2026, which will stabilize its orbit for the next $\sim$ 200 years (gold/orange).
These orbital changes increasingly subject the Centaur's nucleus to more insolation which may drive volatile release.
In Section \ref{sec:therm} we propose a model that explains 450P's activity at each numbered point in this diagram.
The green diamond marker indicates the prediscovery data, the orange hexagons are for data from the literature, and the cyan circular markers are from this work. 
}
\label{fig:orbit}
\end{figure}

One such \ajump\ Centaur is 450P/LONEOS (originally P/2004 A1 (LONEOS), hereafter 450P), which was discovered in 2003 with a conspicuous dust coma at a heliocentric distance of $R_H$ = 5.54 au (\citealt{skiff-2004}).
Orbital integration of 450P's astrometry revealed that it had a close approach to Saturn in 1992 that changed its long-term trans-Saturnian orbit into one with a perihelion closer to Jupiter (Figure \ref{fig:orbit}; \citealt{hahn_2006, lilly-2024ApJ}).
Forward modeling predicts that it will undergo a moderately close encounter (0.5 au) with Jupiter in July 2026 that will stabilize its orbit for the next $\sim$ 200 years.
The combination of 450P's well-constrained orbital history and its recent \ajump\ provides the astronomical community with an almost ideal observational target for investigating fundamental knowledge gaps in our understanding of Centaurs, their links to the TNOs, and how some Centaurs experience new thermal environments en route to becoming JFCs.

We conducted a multi-wavelength observational campaign to measure 450P's activity and composition after its 1992 $a$-jump interaction with Saturn and before its 2026 close encounter with Jupiter. 
Observational details for Gemini-N observations obtained between 2019 and 2024, along with a single JWST dataset acquired in 2023, are presented in Section~\ref{sec:obs}.
Section \ref{sec:results} presents the observational results and
Section \ref{sec:discuss} discusses how these results constrain models of its nucleus and dust and gas in its coma. 
We summarize our key findings in Section \ref{sec:conclusions}.

\bigskip
\section{Observations and Reductions} \label{sec:obs}

We used two observatories to characterize the Centaur's nucleus and coma. 
The Gemini-N 8.1-m telescope on Maunakea, Hawaii was used to recover 450P, measure its nucleus size and  color, and then document its meager dust coma development beyond $\sim$ 7 au. 
The JWST was used for a single observation ``snapshot" in infrared a few weeks after a Gemini-N observation to measure abundances of the suspected dominant volatiles H$_2$O, \cotwo{}, and CO \citep{Mandt2025}, as well as study the morphologies of the dust and gas in the coma.
Figure \ref{fig:orbit} shows the locations in 450P's orbit when the observations were made. 
These are described in more detail in the following two sub-sections.

\subsection{Gemini-N GMOS}

\begin{figure}[h!]
    \centering
    \includegraphics[width=0.32\textwidth]{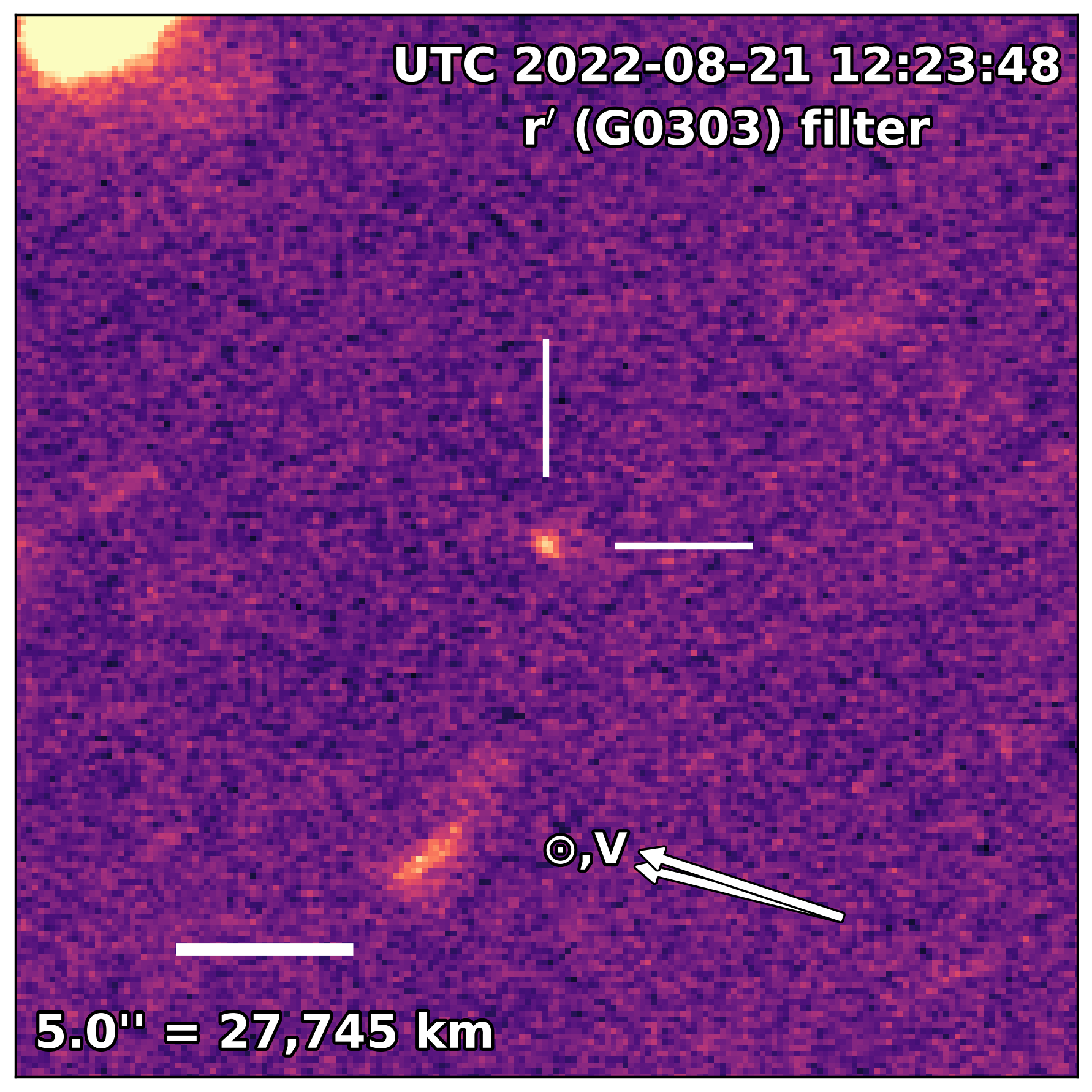}
    \includegraphics[width=0.32\textwidth]{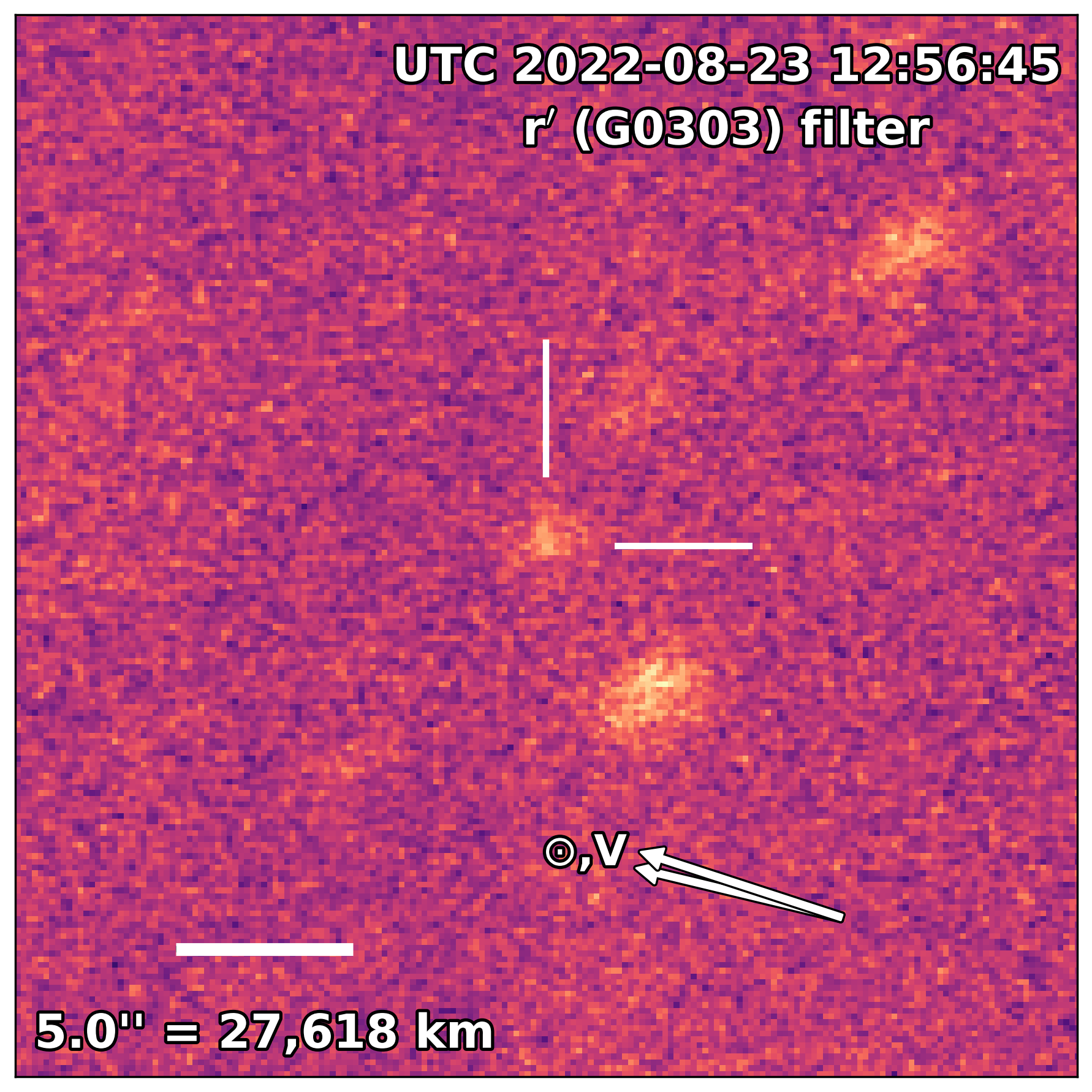}
    \includegraphics[width=0.32\textwidth]{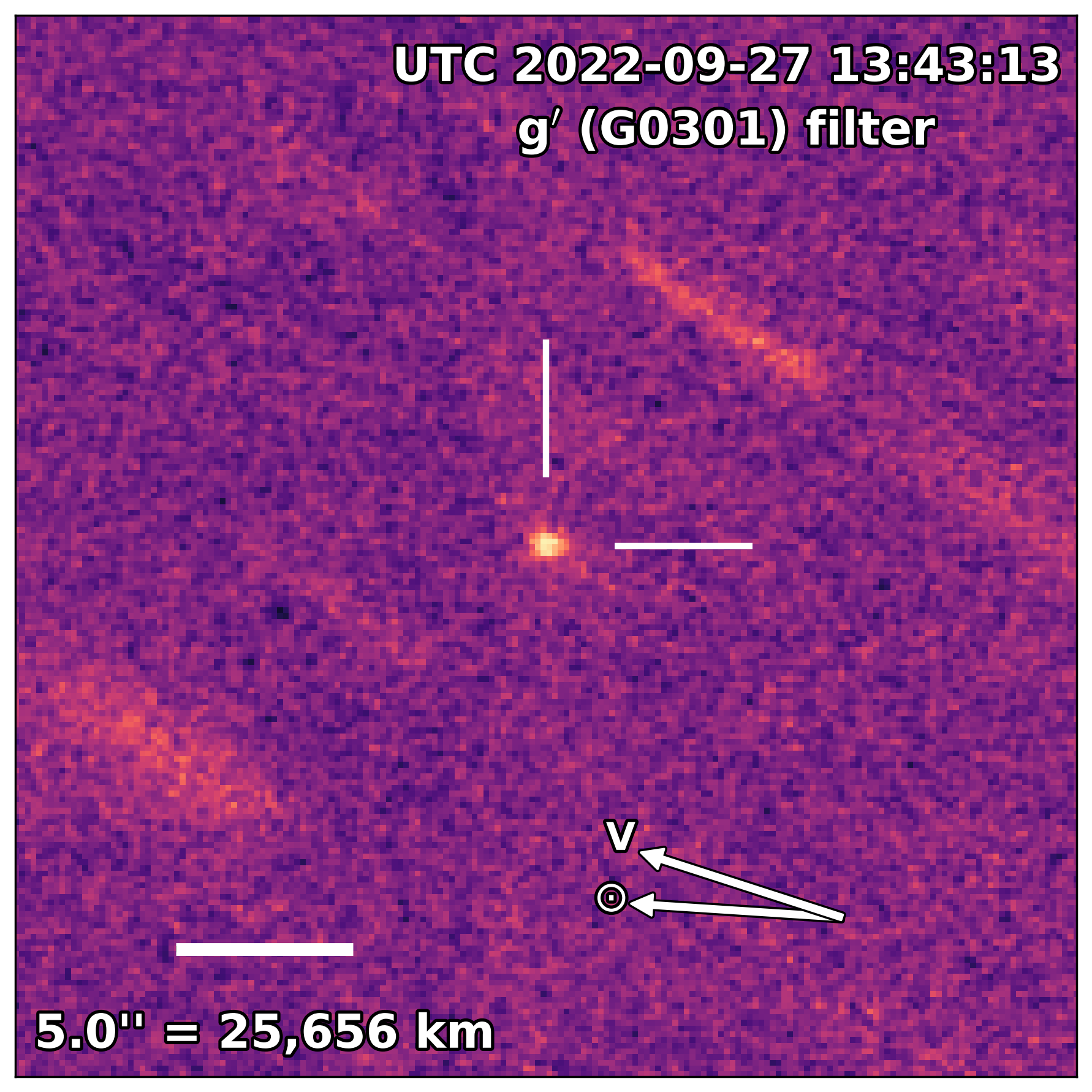} \\
    
    \includegraphics[width=0.32\textwidth]{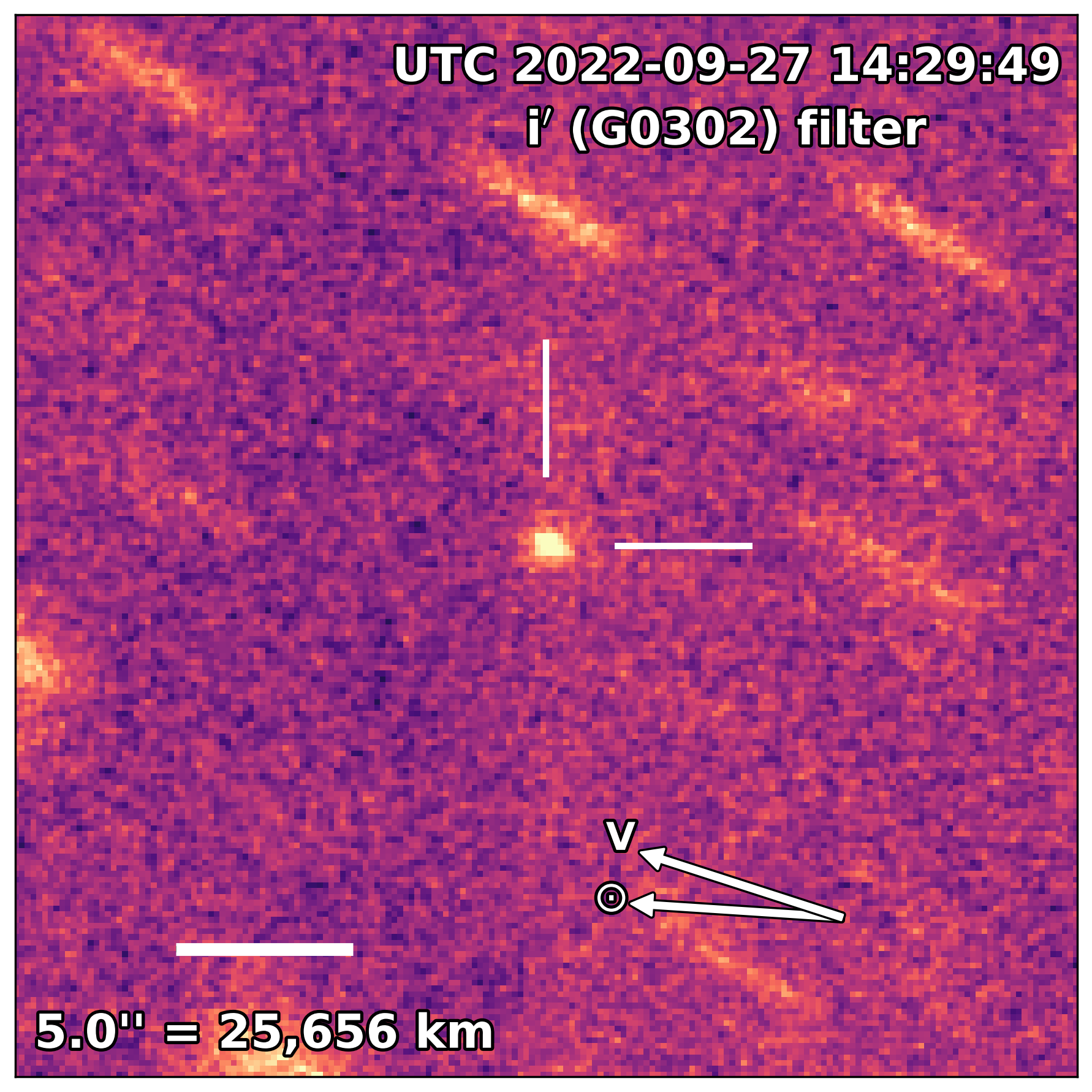}    
    \includegraphics[width=0.32\textwidth]{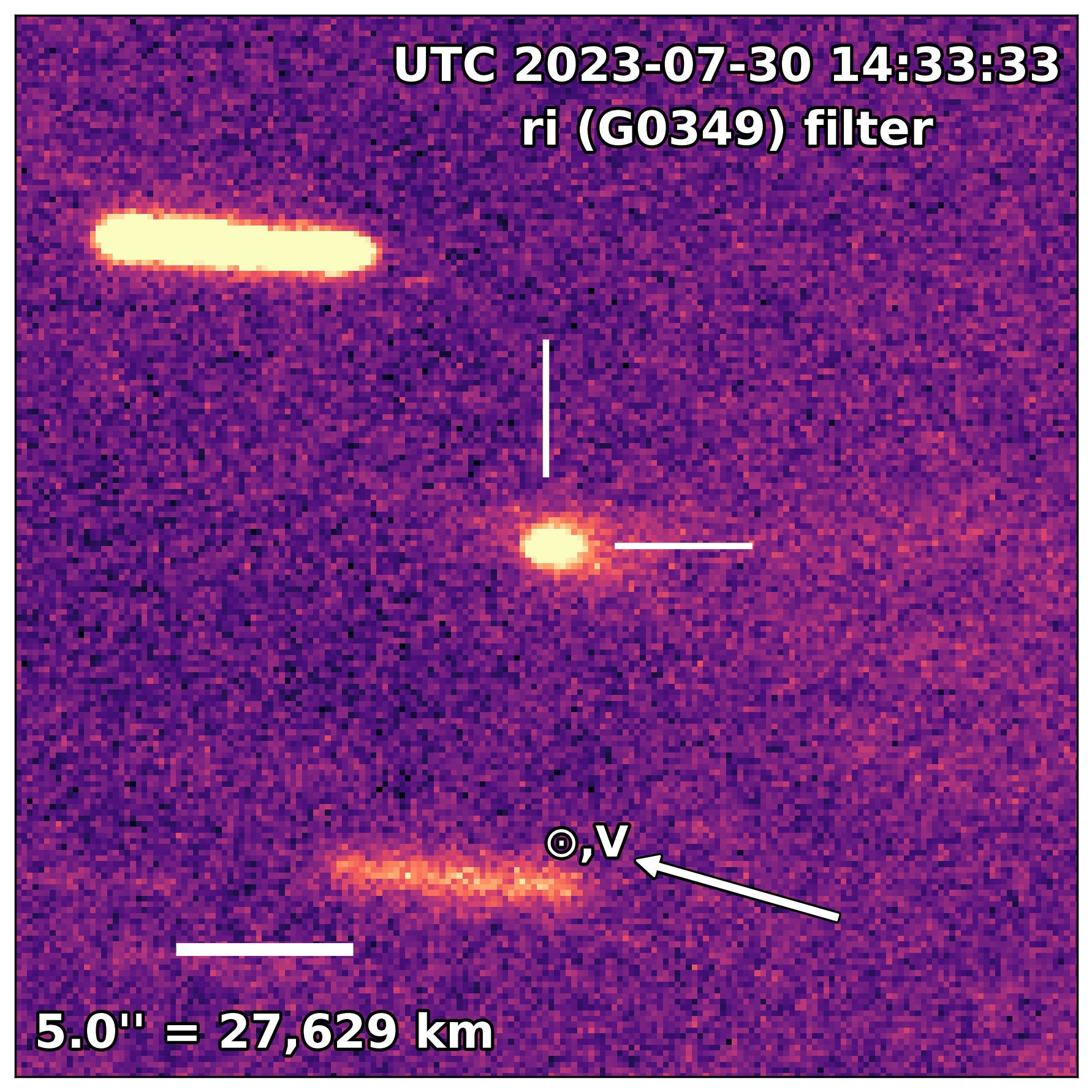}
    \includegraphics[width=0.32\textwidth]{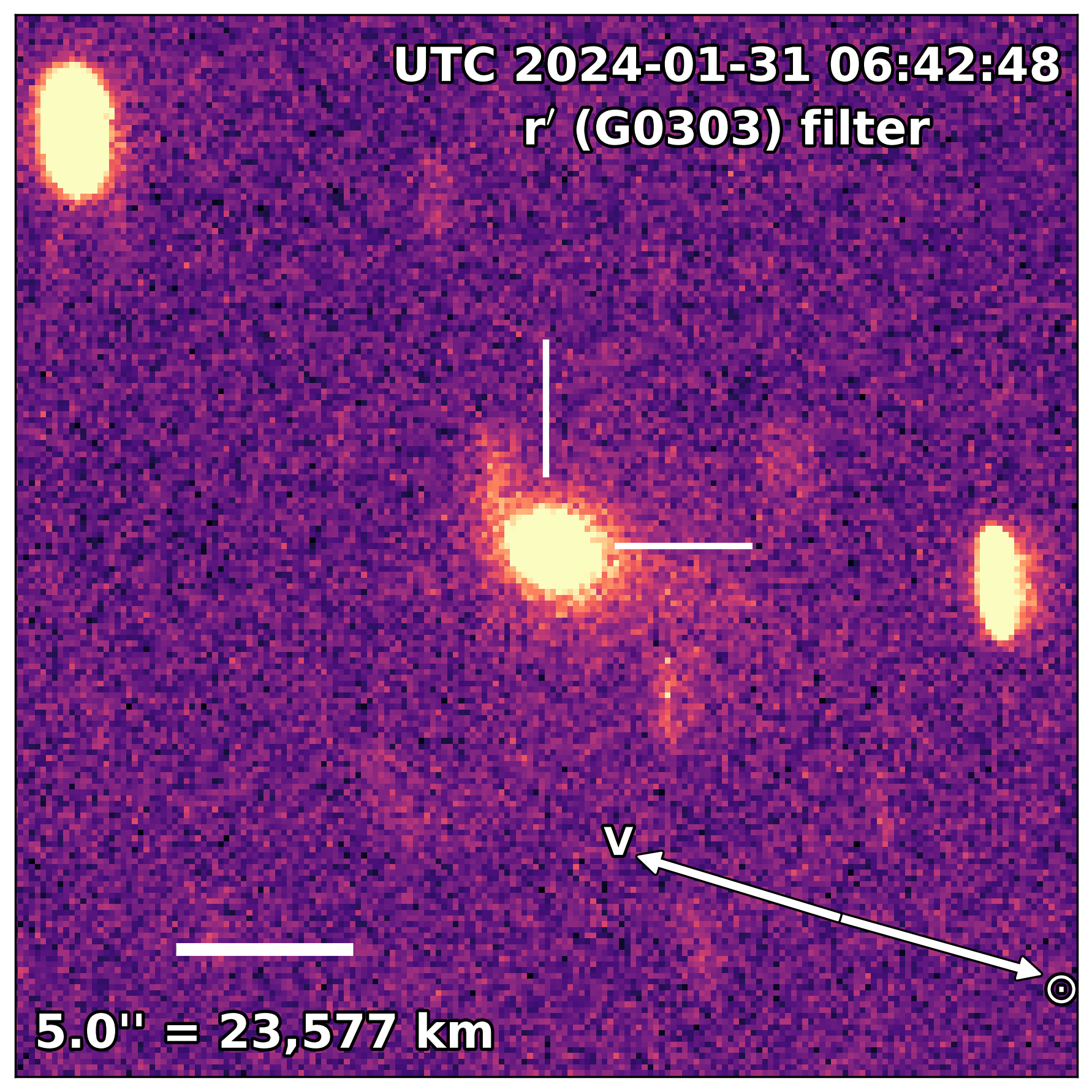}
    
    \caption{Gemini-N GMOS images of 450P. 
    The top-left and top-middle panels display the recovery images when the Centaur appeared inactive and which were used to estimate the nucleus's radius. 
    The top-right and bottom-left panels display the g$'$ and i$'$ images, respectively, when the Centaur also appeared inactive and which were used to measure the nucleus's surface color.  
    The bottom-middle and bottom-right panels display the images where 450P first exhibited a small elongated dust coma. 
    Each panel spans 30$''$ x 30$''$, includes an image scale bar and are oriented with J2000 equatorial North up and East to the left. 
    The skyplane projected directions of the Sun and 450P's velocity are indicated by the white arrows (the appearance of a single white arrow in the bottom-middle panel is due to the alignment of the Sun and 450P velocities).
    }
    \label{fig:gemini-obs}
\end{figure}

As a relatively close-in and active Centaur, 450P was considered by many to be a high priority target for the upcoming JWST. Unfortunately, 450P had not seen for more than 10 years since its 2007 Spitzer observations \citep{fernandez-2006sptz.prop30908F, fernandez_2013} due to its faintness and a poorly constrained orbit. 
The ephemeris reported by the JPL Horizons system had 3-$\sigma$ position uncertainties of several tens of arcseconds, which were too large for the blind pointing and alignment of the NIRSpec IFU's $3\arcsec\times3\arcsec$ field of view (FOV). 
Thus, in 2019 we began a search to recover 450P with the 8.1-m Gemini-N telescope and the Gemini Multi-Object Spectrograph (GMOS) in imaging mode \citep{hook-2004PASP-gmos}.
After three unsuccessful attempts, we recovered 450P on UTC 2022-08-21 and confirmed it two nights later.
Astrometry information was submitted to the Minor Planet Center (MPC) to help significantly decrease 450P's ephemeris uncertainties \citep{450P-recovery-2022CBET.5178....1R}.
In total, we obtained five nights of optical imaging data over the next 1.5 years using a variety of broadband filters with Gemini-N GMOS (see Table \ref{tab:obs} and Figure \ref{fig:gemini-obs}). 
The GMOS filters used are similar to the Sloan Digital Sky Survey (SDSS) filters \citep{fukugita-1996AJ-sdss-filters} aside from the ri filter (G0349), which has a top-hat transmission curve covering both the r$'$ and i$'$ bandpasses.
The detector pixels were binned 2 x 2 for all observations, resulting in an $\sim$~0$''$.16 pixel scale.

\begin{figure}[h!]
    \centering
    \includegraphics[width=0.45\textwidth]{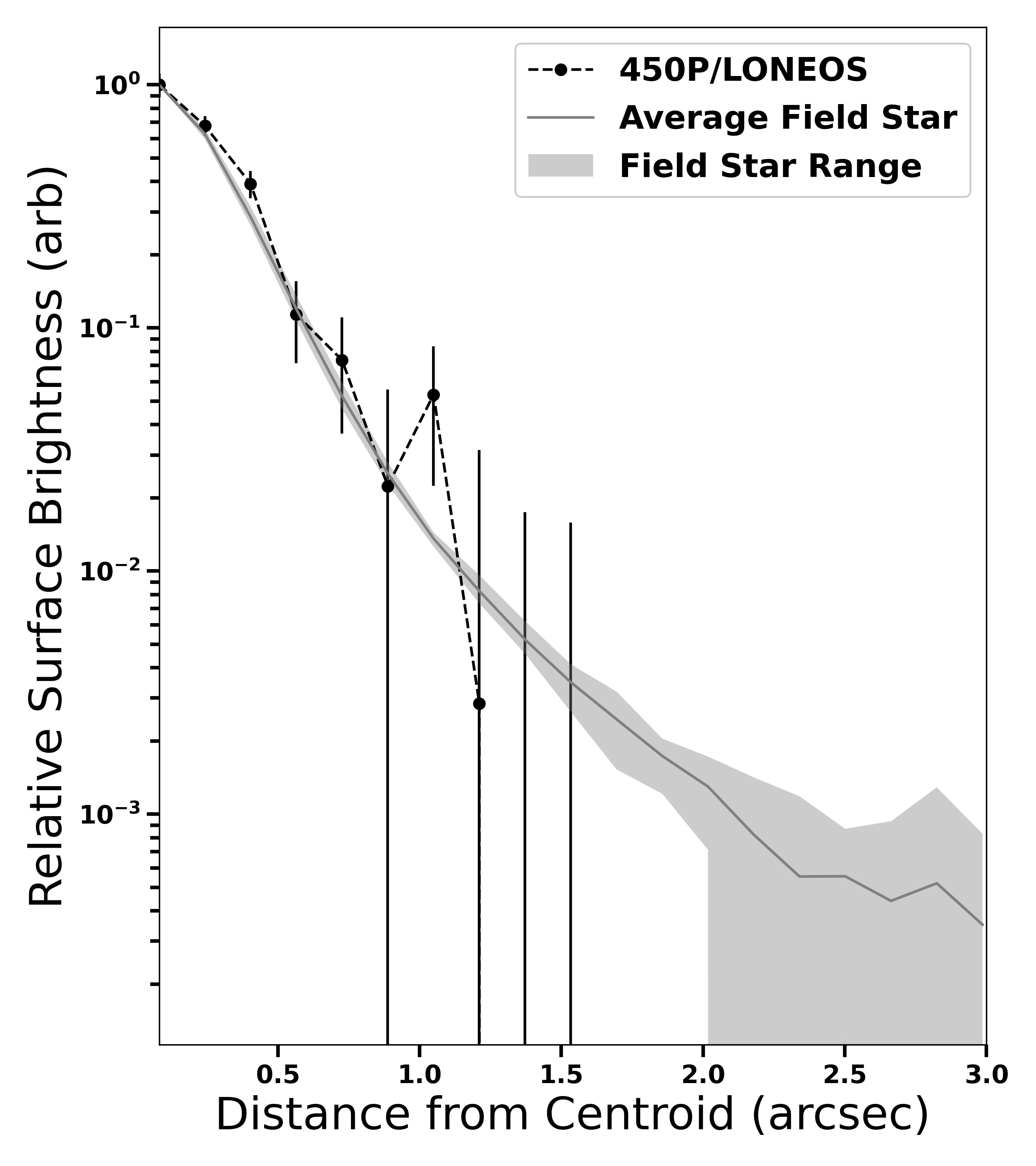}~
    \includegraphics[width=0.45\textwidth]{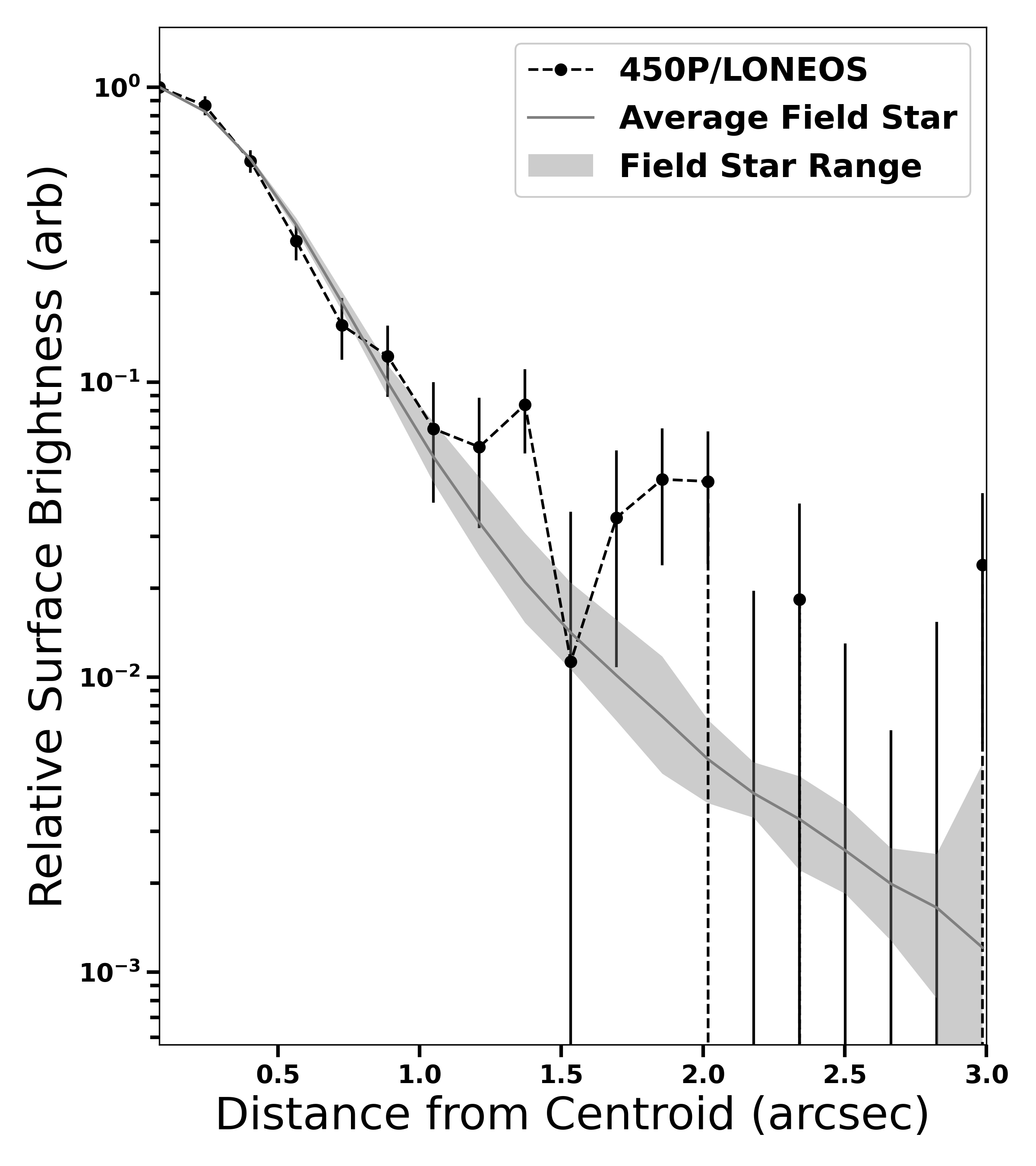}~
    \caption{
    Normalized radial surface brightness profiles of the Gemini-N 450P images on 2022-08-21 (r$'$; left panel) and 2022-09-27 (g$'$; right panel). Surrounding field star profiles are overplotted with a gray shaded region representing their range of scatter. 
    450P's profiles are similar to those of the field stars, which is consistent with little-to-no dust coma present.
    }
    \label{fig:stellar-profile-comparison}
\end{figure}

The recovery images displayed a point-source object (see Figure \ref{fig:stellar-profile-comparison}) consistent with the predicted positions and skyplane motion but at apparent visible magnitudes (SDSS r$'$) approximately two magnitudes fainter than predictions based on nucleus size estimates made during its last apparition \citep{hahn_2006}.
Thus, we acquired g$'$ and i$'$ images on 2022-09-27 with the goal of providing the first color measurement of the presumed bare-nucleus.

Bias subtraction, flat-field application, and mosaicking of the GMOS's three CCDs into a single image were completed using the Gemini DRAGONS software \citep{labrie-2019-dragons}. 
Individual images were then processed using our Python-based GMOS reduction and calibration pipeline, which includes (1) cosmic-ray removal performed using the LACosmic technique \citep{van-dokkum-2001} as implemented in \texttt{ccdproc} \citep{craig-2017}, (2) stacking of each filter's individual images in 450P's non-sidereal and sidereal frames, and (3) a Pan-STARRS-based calibration using field stars from the central GMOS CCD (which included 450P) stacked in the sidereal frame.

The magnitudes used for image calibrations are from the Pan-STARRS 1 (PS1) data release 2 (DR2) MeanObject tables reported in the AB magnitude system \citep{magnier-2013ApJ=Pan-STARRS}, where the PS1 magnitudes were converted to the SDSS system using the equations of \cite{tonry_2012}.
Corresponding SDSS AB magnitude photometry upper-limits and measurements are included in Table \ref{tab:obs}.

\begin{deluxetable*}{cccccccccc}[h!] \label{tab:obs}
\tablecaption{Gemini Observing Details and Photometry}
\tablecolumns{10}
\tablewidth{0pt}
\tablehead{
\colhead{UTC Date} &
\colhead{UTC Time\tablenotemark{a}} &
\colhead{$R_H$} &
\colhead{$\Delta$} &
\colhead{$\alpha$\tablenotemark{b}} &
\colhead{Seeing} &
\colhead{Filter\tablenotemark{c}} &
\colhead{Total Exp. Time} &
\colhead{Airmass\tablenotemark{d}} &
\colhead{$m$}\\      % Second line with the units of quantities.
\colhead{(YYYY-MM-DD)} & 
\colhead{(d)} &
\colhead{(au)} &
\colhead{(au)} &
\colhead{$(^{\circ})$} &
\colhead{(arcsec)} & 
\colhead{} & 
\colhead{(seconds)} &
\colhead{} &
\colhead{(mag)} 
}

\startdata
2019-07-05 & 09:18:17 & 9.663 & 9.494 & 6.003 & 0$''$.89 & r$'$ (G0303) & 1400 & 1.16 & $>$ 25.0 \\ 
2021-07-14 & 14:09:18 & 8.622 & 8.711 & 6.700 & 1$''$.00 & r$'$ (G0303) & 800 & 1.40 & $>$ 25.4 \\
2021-10-11 & 09:57:47 & 8.471 & 7.505 & 1.816 & 0$''$.91 & r$'$ (G0303) & 960 & 1.10 & $>$ 25.6 \\
2022-08-21 & 12:23:48 & 7.897 & 7.642 & 7.228  & $0\arcsec.57$ & r$'$ (G0303) & 1800 & 1.38 & 25.08 $\pm$ 0.07  \\
2022-08-23 & 12:56:45 & 7.894 & 7.607 & 7.178 & $1\arcsec.24$ & r$'$ (G0303) & 2080 & 1.24 & 24.85 $\pm$ 0.07 \\
2022-09-27 & 13:43:13 & 7.827 & 7.068 & 5.052 & $0\arcsec.88$ & g$'$ (G0301) & 2340 & 1.08 & 24.94 $\pm$ 0.07 \\
" %  
& 14:29:49 & 7.826 & 7.067 & 5.048 & $0\arcsec.73$ & i$'$ (G0302) & 1560 & 1.15 & 23.79 $\pm$ 0.05 \\
2023-07-30 & 14:33:33 & 7.225 & 7.609 & 7.265 & $0\arcsec.68$ & ri (G0349) & 1320 & 1.48 &  22.60 $\pm$ 0.02\tablenotemark{e}  \\
2024-01-31 & 06:42:48 & 6.857 & 6.505 & 7.869 & $0\arcsec.54$ & r$'$ (G0303) & 1600 & 1.09 &  21.52 $\pm$ 0.01\tablenotemark{e}  \\
\enddata

\tablenotetext{a}{UTC at start of image sequence.}
\tablenotetext{b}{The solar phase angle at the time of the observation.}
\tablenotetext{c}{For specifics of the GMOS filters see \cite{gemini:gmos-filters}.}
\tablenotetext{d}{Mean airmass of 450P during the exposure.}
\tablenotetext{e}{Measured using a circular aperture with projected skyplane radius of 10,000 km, 1$''$.81 and 2$''$.12, respectively for the dates of UTC 2023-07-30 and 2024-01-31. Other magnitude measurements reported in the table are for point-sources.}

\end{deluxetable*}

\subsection{JWST NIRSpec}

Infrared spectra of 450P were acquired on UTC 2023-09-03 between 10:18:20 -- 15:17:10 using the JWST NIRSpec instrument \citep{jakobsen-2022, boker-2023nirspec} in the Integral Field Unit (IFU) mode when the Centaur was at $R_H =$ 7.16 au, the telescope--450P distance was 7.01 au, and a solar phase angle was 8.15$^{\circ}$. 
Spectra were obtained using the JPL Horizons' predicted ephemeris positions and the JWST 4-point dither pattern with offsets of $\sim$ 0\farcs4 between individual exposures resulting in an effective on-target exposure time of 2188.333 s.
The PRISM/CLEAR disperser-filter combination was used, providing wavelength coverage from 0.6--5.3 $\mu$m with a spectral resolving power ranging from $R \sim 30$--300 \citep{boker-2023nirspec}.
This wavelength range covers the gas emissions from the rovibrational bands of H$_2$O, \cotwo{}, and CO as well as absorption features of water-ice.
The IFU FOV covers a 3$''$ × 3$''$ square region with spaxels that are 0\farcs1 $\times$ 0\farcs1.

An additional set of observations using the same configuration were acquired at an offset of 60 arcseconds from 450P's predicted ephemeris position at a position angle of +90$^{\circ}$ with respect to the Sun's skyplane projected direction. 
These observations were used to estimate and remove the background from the on-target observations.

The raw uncalibrated data for each dither position and offset were downloaded from the MAST database using the \texttt{astroquery} Python package \citep{ginsburg-2019AJ-astroquery}.
The data can be found at \dataset[doi: 10.17909/2vtm-st63]{https://doi.org/10.17909/2vtm-st63}.
Each individual datacube was then processed with JWST pipeline version v1.19.2 \citep{bushouse_2025_zndo..16280965B}.
The Calibration Reference Data System (CRDS) context file {\it jwst}\_1413.{\it pmap} was used for the data reductions.
The CRDS context file used can be found at: \url{https://jwst-crds.stsci.edu/}.

A median-combined datacube was generated using the 4 individual datacubes from the on-target observations. 
The median-combined datacube was constructed by aligning the slices from the four individual datacubes in 450P’s non-sidereal frame and computing the median value at each spaxel.
A similar procedure was applied to the offset observations.
Background subtraction was performed by subtracting the median-combined offset datacube from the on-target median-combined datacube on a slice-by-slice basis.

\begin{figure}[h!]
    \centering
    \includegraphics[width=0.8\textwidth]{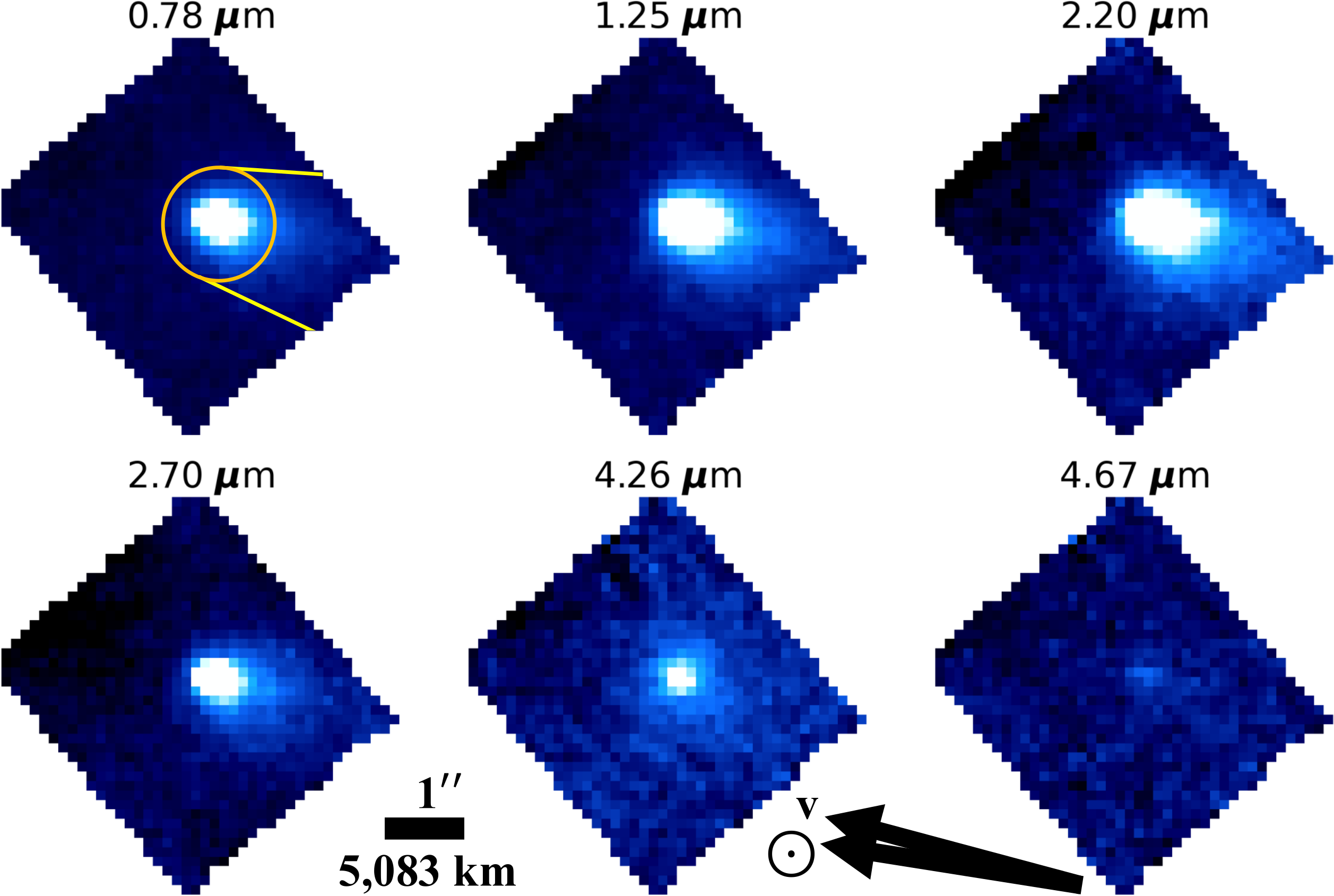}
    
    \caption{Wavelength-integrated image panels of 450P extracted from JWST NIRSpec IFU datacube at different bandpasses (the effective wavelength is listed directly above each panel). 
    The panels use the same color scale, ranging from black through blue to white, to indicate increasing surface brightness.
    The panels are shown with J2000 equatorial north up and east to the left; the sky-plane–projected directions of the Sun and 450P’s velocity vectors are indicated by black arrows.
    The image scale bar is the same for each panel.
    A faint surface-brightness enhancement is seen extending toward the west–southwest of each panel which is consistent with a coma and tail formation indicated in the top-left panel by an orange circle and yellow lines, respectively.
    Although the bottom-left panel’s bandpass includes the \water{} emission band at 2.70~$\mu$m and the bottom right includes the 4.67~$\mu$m band of CO, no emission feature was detected for either in the extracted spectrum (see Figure~\ref{fig:jwst-spectrum}) and, thus, the observed surface brightness at these wavelengths is dominated by reflected flux from the nucleus and dust in the coma.
    The image at 4.26 $\mu$m contains contributions from both dust and a prominent \cotwo{} emission feature which dominates the coma morphology and appears more circular than the coma at other wavelengths that are dominated by dust continuum.
    }
    \label{fig:jwst-obs}
\end{figure}

Wavelength-integrated image panels were constructed by summing individual datacube slices over wavelength ranges corresponding to standard broadband astronomical filters and the emission bandpasses of \water{}, \cotwo{}, and CO (see Figure~\ref{fig:jwst-obs}).
The specific bandpasses and their effective wavelengths are as follows: 
(top left) SDSS i$'$ filter at 0.78 $\mu$m, 
(top middle) J-band filter at 1.25 $\mu$m, 
(top right) K-band filter at 2.20 $\mu$m, 
(bottom left) \water{} $\nu_3 + \nu_1$ 
rovibrational emission complex at 2.70 $\mu$m, 
(bottom middle) \cotwo{} $\nu_3$ emission band at 4.26 $\mu$m, and 
(bottom right) the CO v=1-0 emission band at 4.67 $\mu$m. 
Prominent coma extensions are visible in the JWST IFU panels.

\begin{figure}[h!]
    \centering
    \includegraphics[width=0.50\textwidth]{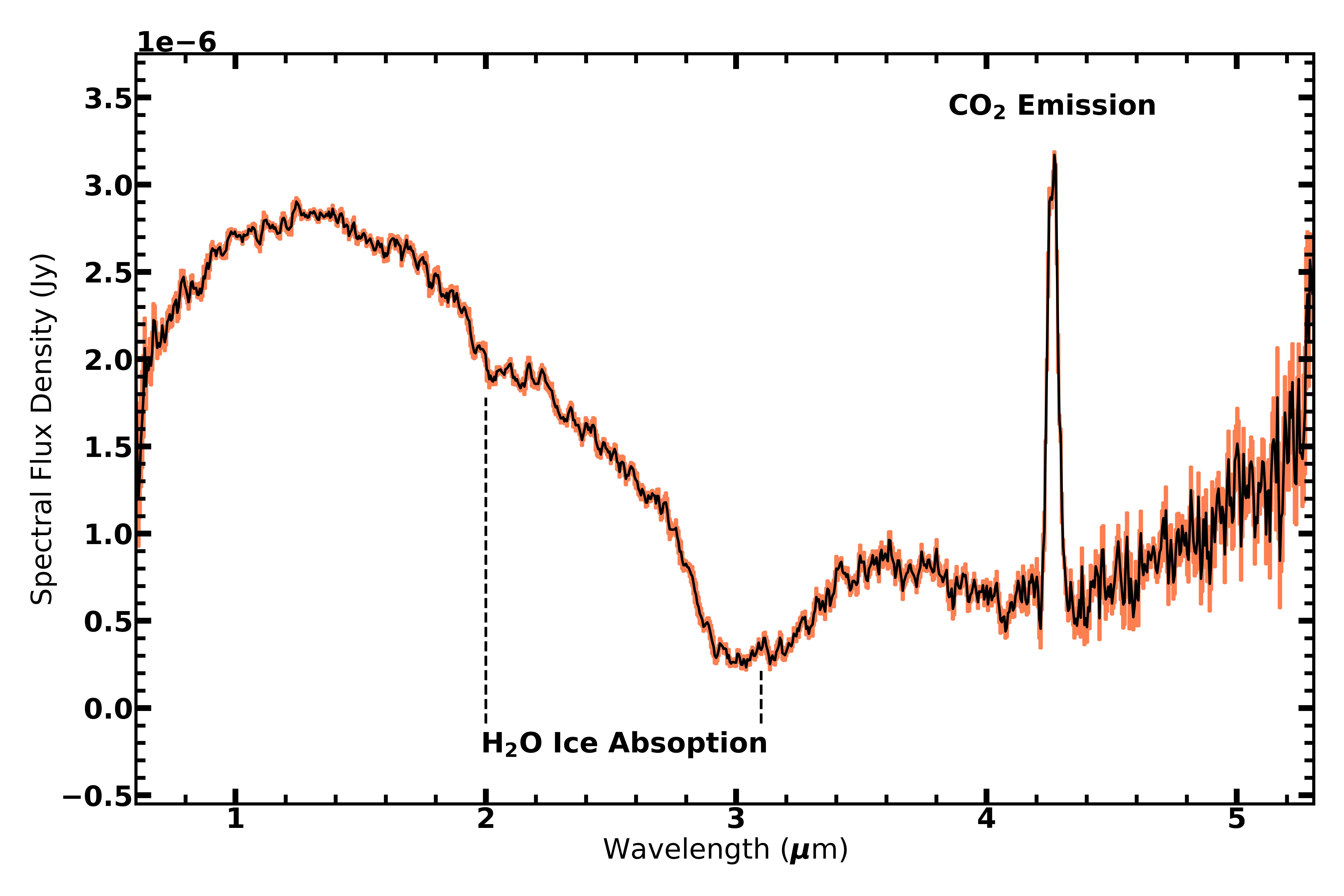}~
    \includegraphics[width=0.5\textwidth]{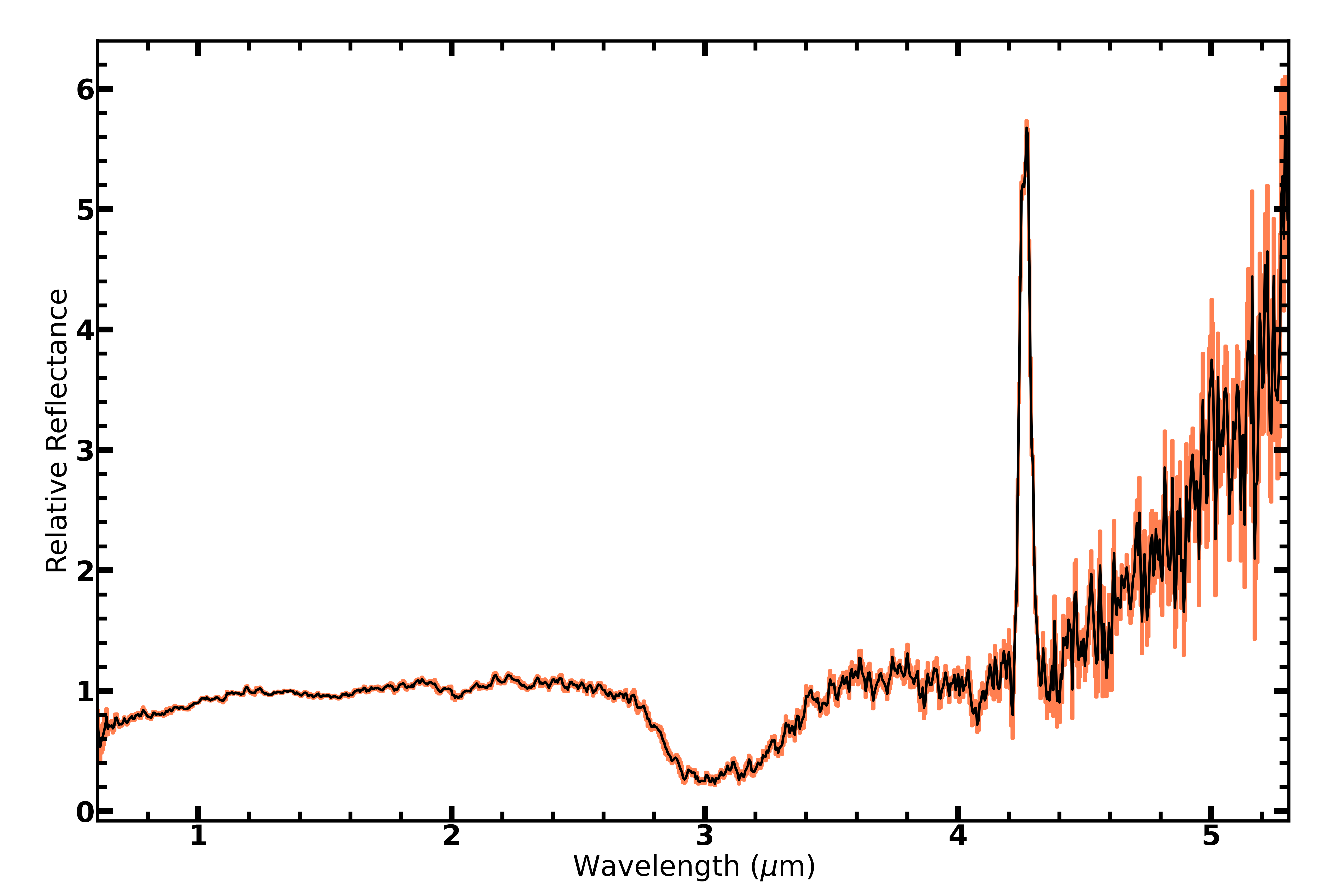}
    
    \caption{Left: 
    The spectrum of Centaur 450P at $R_H$=7.16~au obtained with JWST NIRSpec on 2023-09-03.
    The black line is the spectrum extracted from the median-combined datacube with the coral-red shaded region indicating the spectrum's 1-$\sigma$ uncertainty.
    The spectrum is dominated by reflected solar radiation, a prominent \cotwo{} emission feature at 4.26 $\mu$m, and likely water ice absorption features at 2.0 and 3.0 $\mu$m.
    No emission lines were detected from \water{} at 2.7 $\mu$m or CO at 4.65 $\mu$m.
    Right: 450P's spectrum after division by the solar analog star SNAP-2 spectrum and normalized to one at 2.6 $\mu$m.
    }
    \label{fig:jwst-spectrum}
\end{figure}

A 1D spectrum covering the full wavelength range was extracted from the median-combined, background subtracted datacube using aperture photometry with a circular aperture of radius 0$''$.45, centered on 450P’s position as determined from the slice centroid.
The skyplane projected radius of the circular aperture at the distance of 450P was 2,541 km.
Figure \ref{fig:jwst-spectrum}, left panel displays a spectrum of 450P. 
The 1-sigma uncertainty was estimated using the local variability of the spectrum through a rolling standard deviation measurement using a window size of seven resolution elements. 
The \cotwo{} emission feature was masked between wavelengths of 4.21 - 4.36 $\mu$m where the uncertainties covering the \cotwo{} emission feature were assumed to follow the statistical behavior of the surrounding wavelengths.
The spectrum is dominated by reflected light, a broad 3-$\mu$m absorption band, and the 4.26~$\mu$m \cotwo{} emission feature.
Figure \ref{fig:jwst-spectrum} right panel displays the relative reflectance spectrum generated by dividing 450P's spectrum by a spectrum of the solar analog star SNAP-2 \citep{1128-2019jwst.prop.1489L} and normalizing the result to one at 2.6 $\mu$m.

\section{Observational Results} \label{sec:results}

\subsection{Gemini-N images and radial profiles}

Figure \ref{fig:gemini-obs} shows the non-sidereal stacked images cropped and centered on 450P over five nights between 2022-08-21 and 2024-01-31. 
Visual inspection shows no extension beyond a point-source in any of the 2022 images and normalized radial surface brightness profiles agree well with those of surrounding field stars (Figure \ref{fig:stellar-profile-comparison}), consistent with the lack of any substantial coma. 
The 450P profiles were derived from the non-sidereal stacked images and the star profiles were derived from the sidereal stacked images. The 2022 images may represent the first observations of the previously active Centaur's bare nucleus, although a very low-level of dust coma cannot be ruled out.

450P displayed a dust coma on the final two nights (UTC 2023-07-30 and 2024-01-31), indicating that activity commenced when the Centaur moved between heliocentric distances of $R_H$ = 7.83 au and 7.23 au. 
The coma shape was asymmetric and elongated along the tailward direction.

\subsection{JWST NIRSpec datacubes: spectra, images, and radial profiles}
\label{sec:irdata}

\begin{figure}[h!]
    \centering
    \includegraphics[height=0.405\textwidth]{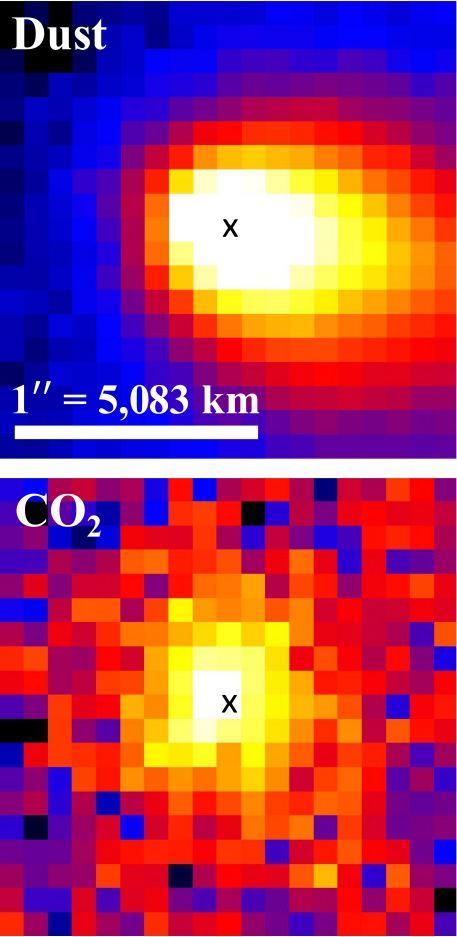}
    \includegraphics[height=0.40\textwidth]{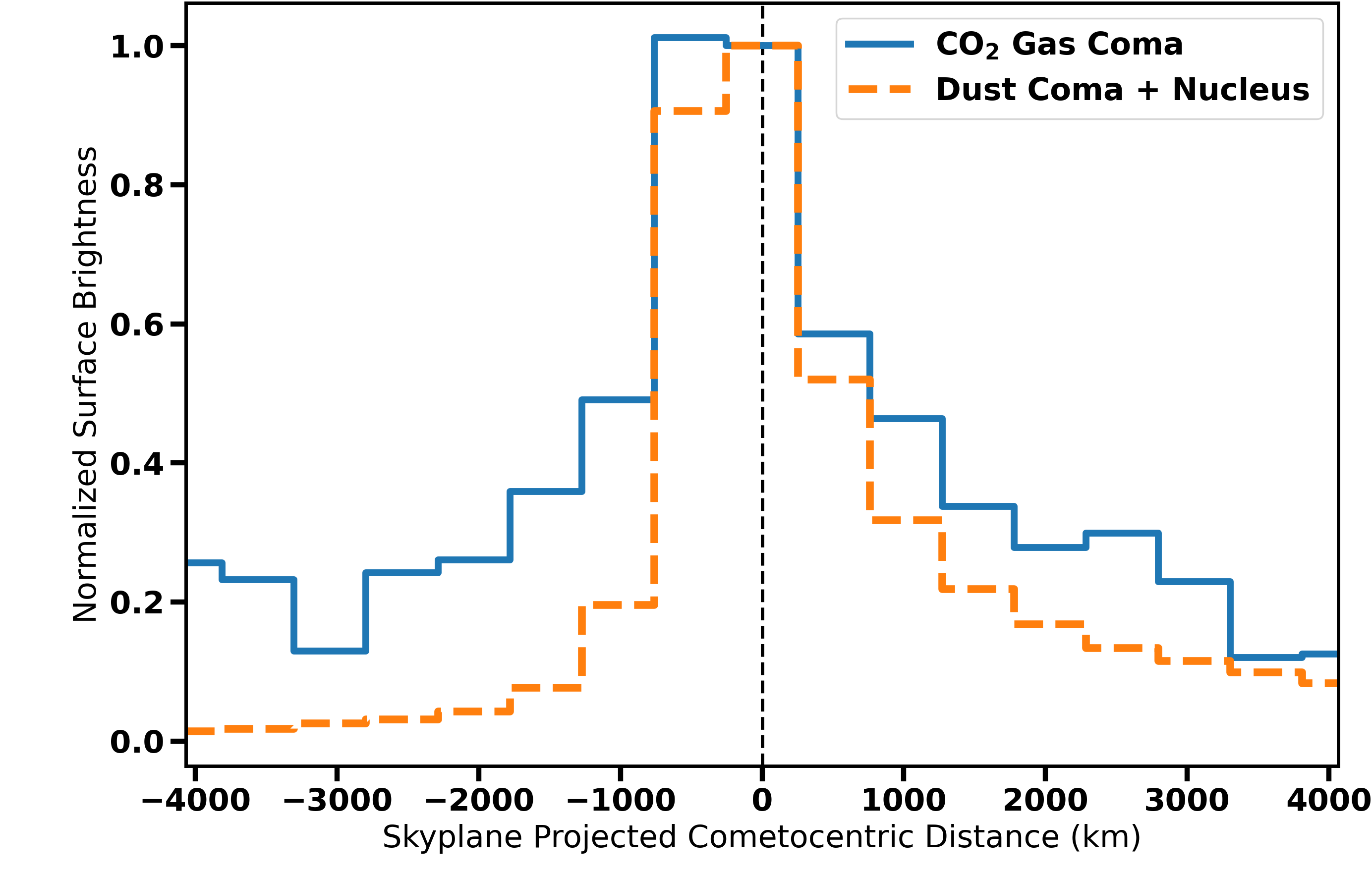}

    \caption{Measurements of the dust and \cotwo{} gas emission in 450P's coma. 
    The left two panels display the dust and \cotwo{} gas surface brightness maps.
    The location of 450P's nucleus is indicated by the ``x" in each image and 
    the panel orientations and skyplane velocity vector directions are the same as in Figure \ref{fig:jwst-obs}.
    A scale bar is included in the dust image.
    The dust image shows elongated structure with a tail while the \cotwo{} is much more symmetric.
    The right panel displays radial surface brightness profiles for \cotwo{} (solid blue line) and dust (dashed orange line) plotted against projected cometocentric distance (sunward corresponds to negative values and anti-sunward/tailward corresponds to positive values of the abscissa). 
    Each radial profile is normalized to 1 in the central pixel and the dust profile also includes the nucleus's contribution.
    For reference, the peak values of each profile before normalization are 227.95 MJy sr$^{-1}$ for the nucleus~$+$~dust and 3.57~MJy~sr$^{-1}$ for the \cotwo{} profiles.
    }
    \label{fig:ccoma-morpho}
\end{figure}

Figure~\ref{fig:jwst-obs} displays JWST NIRSpec IFU panels of 450P at selected wavelength regions. 
Figure \ref{fig:jwst-spectrum} shows the 1D extracted spectrum with clear detection of \cotwo{} emission and likely H$_2$O absorption bands.
A \cotwo{}-gas-only surface brightness image (at 4.26 $\mu$m) was generated by removing the dust and nucleus contributions to the datacube slices covering the \cotwo{} emission band using the median-combined datacube.
The datacube's FOV was cropped to approximately 2$'' \times$ 2$''$ and centered on the nucleus position. 
Next, the datacube was separated into three wavelength regions: 
(4.131–4.216~$\mu$m), 
(4.221–4.306~$\mu$m), 
and (4.316–4.402~$\mu$m), 
corresponding to spectral regions shortward of, covering, and longward of the \cotwo{} emission band, respectively.
Each wavelength region was then summed on a spaxel-by-spaxel basis, resulting in three wavelength-integrated 2D maps of the integrated surface brightness.
An effective nucleus and dust coma surface brightness map was generated for a wavelength of 4.26 $\mu$m, the center of the \cotwo{} emission band, by performing a linear fit to the pixels of the shortward and longward maps, where the linear fit was used to calculate the map at 4.26 $\mu$m.
We note that in this wavelength region the spectrum's shape is likely not linear, but for this initial study a linear fit was assumed sufficient over the small wavelength interval.
The nucleus and dust coma map at 4.26 $\mu$m was then subtracted from the surface brightness map covering the \cotwo{} emission.
The resulting CO$_2$-only surface brightness distribution is shown in the lower-left panel of Figure \ref{fig:ccoma-morpho}.
Note that this CO$_2$-only map differs from the lower-middle panel in Figure~\ref{fig:jwst-obs}, which also includes contributions from the nucleus and dust in the coma.

We generated a surface brightness map of dust in the coma by summing the median-combined, cropped datacube over the wavelength interval 0.6575–2.0025~$\mu$m.
This map (see Figure \ref{fig:ccoma-morpho} upper left) includes contributions of both the nucleus and coma dust, as no nucleus-only data are available to remove the nucleus's contribution.
The extended surface brightness indicates that dust in the coma dominates the detected flux and we assume to first order that the nuclear point-source contribution is negligible.

As seen in the \cotwo{} and dust surface brightness maps, the morphologies show considerable differences.
The \cotwo{} surface brightness appears to be azimuthally symmetric while the dust displays asymmetry with elongation along the tailward direction.
These differences are highlighted when comparing radial surface brightness profiles along the direction of 450P's projected Sun-to-nucleus vector (Figure \ref{fig:ccoma-morpho} right panel).
The \cotwo{} morphology is consistent with emission of molecules from a cometographic location near the sub-solar point with a 3D cone shape similar to the CO gas morphology seen in the coma of Centaur 29P/Schwassmann-Wachmann 1 \citep{faggi-2024NatAs}.
The dust surface brightness displays a compact coma and signs of tail formation under the influence of solar radiation pressure.

\begin{figure}[h!]
    \centering
    \includegraphics[width=0.65\textwidth]{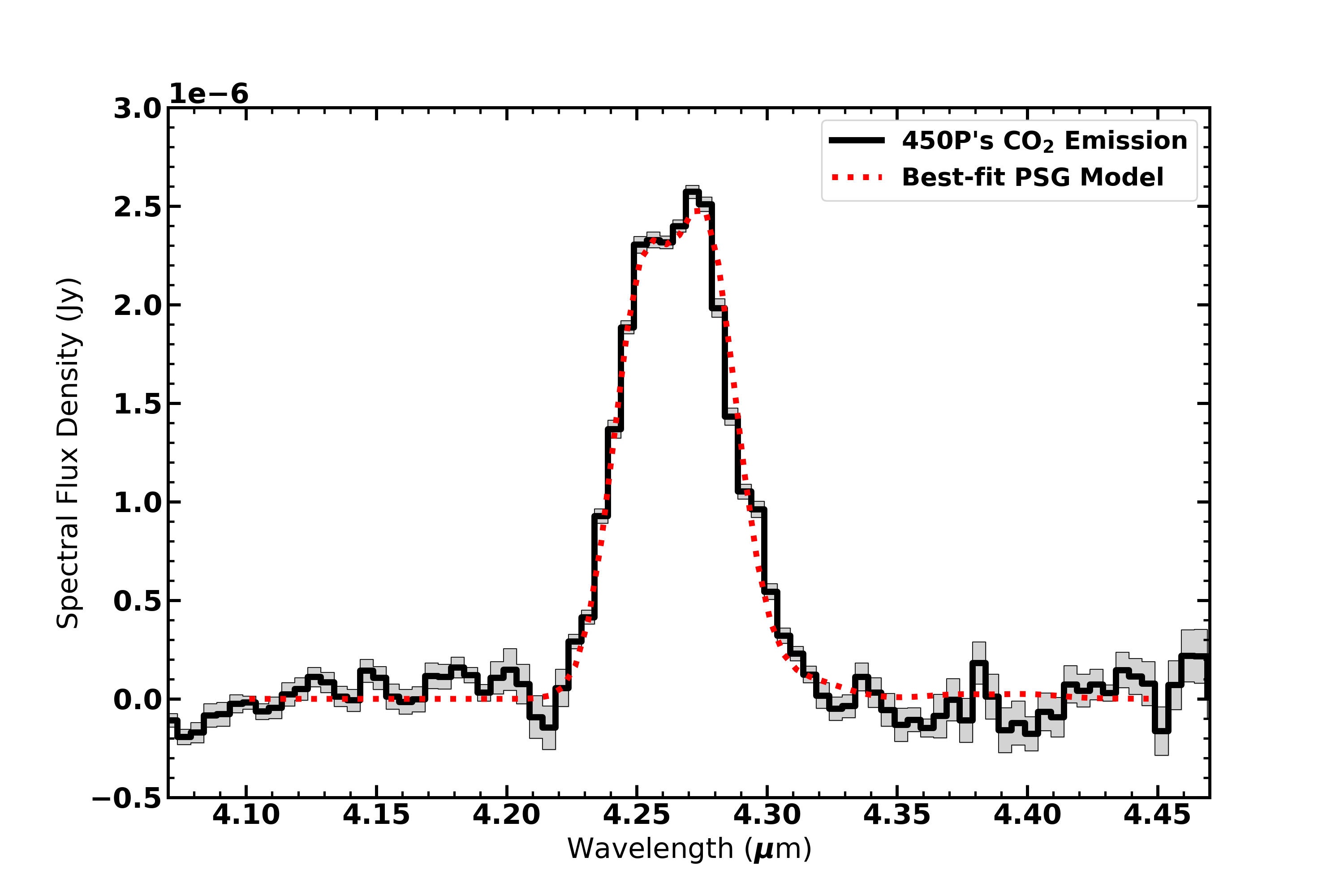}
    
    \caption{A close-up of the continuum subtracted JWST spectrum of 450P centered on the CO$_2$ emission band (black line) with the best-fit model spectrum of the CO$_2$ $\nu_3$ fluorescence emission band overplotted in a red dashed line using the Planetary Spectrum Generator.  
    The model is consistent with $Q_{\mathrm{CO2}}$ = \cotwoproduction{}~molec.~s$^{-1}$ and $T_\mathrm{rot}$~=~60~$\pm$~1~K. 
    The associated 1-sigma uncertainties for the spectrum are shown with the gray shaded regions. 
    }
    \label{fig:co2-spec}
\end{figure}

\begin{figure}[h!]
    \centering
    \includegraphics[width=0.48\textwidth]{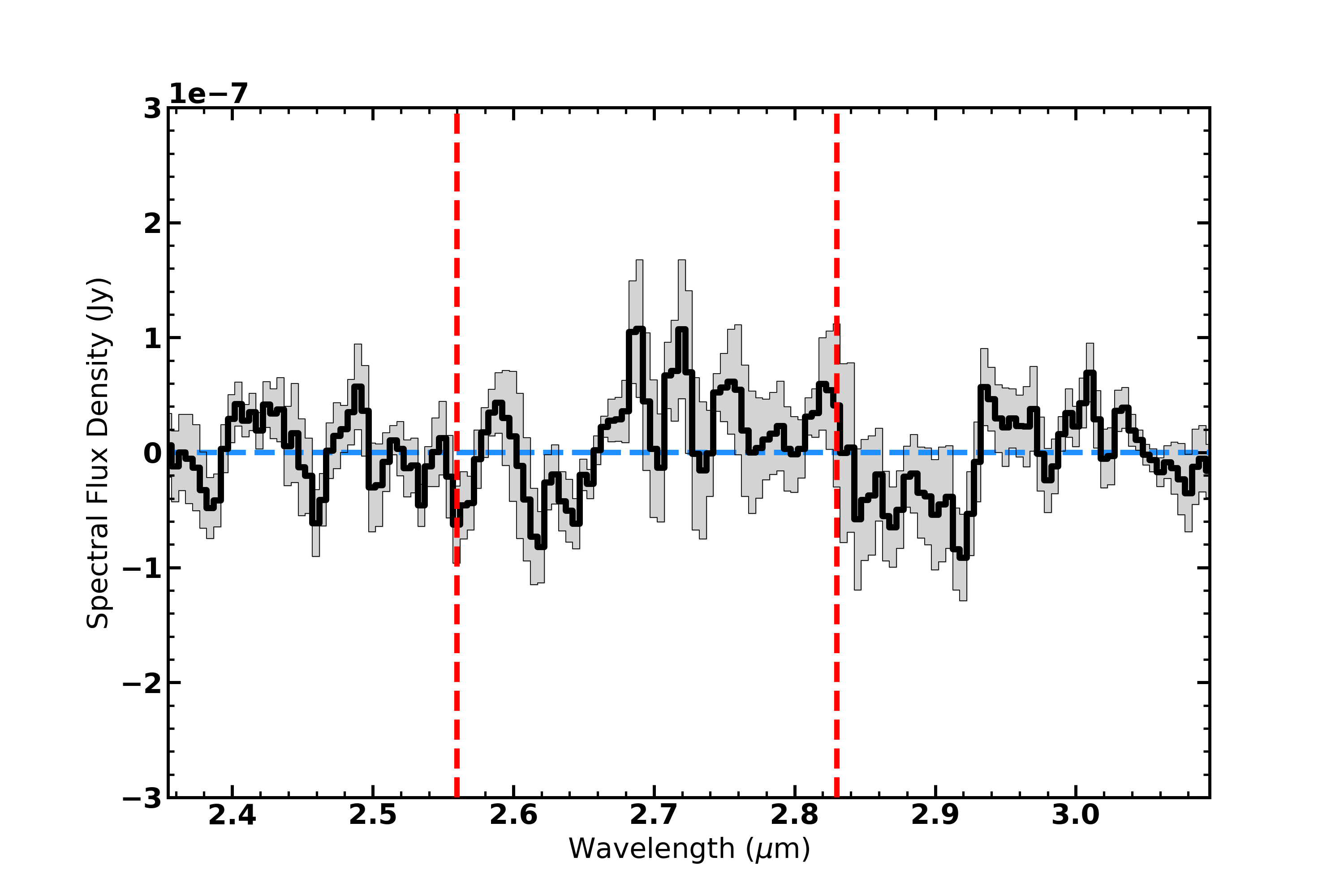}
    \includegraphics[width=0.48\textwidth]{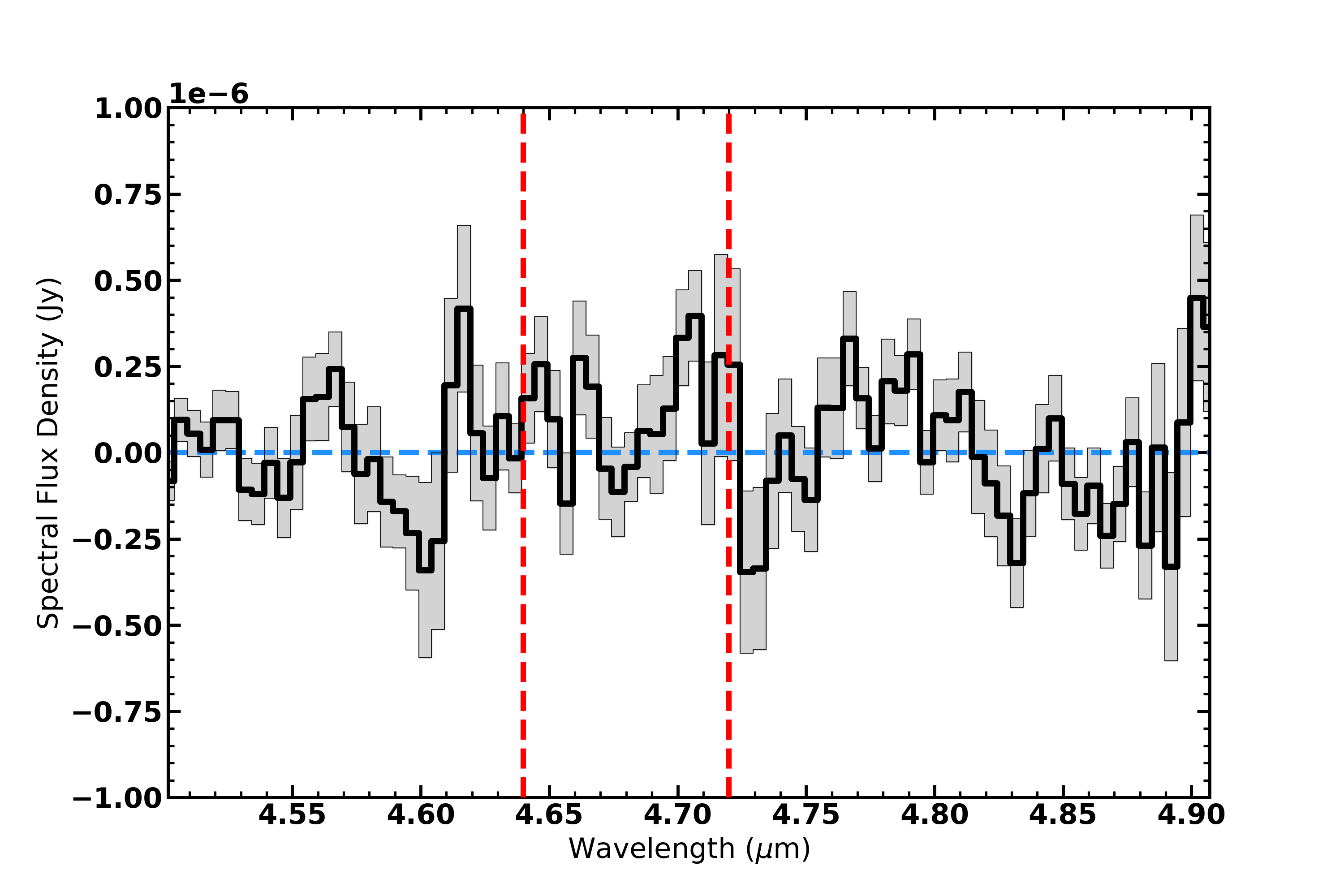}

    \caption{The continuum subtracted JWST spectrum (black line) for the spectral ranges where  \water{} (left panel) and CO (right panel) emission bands would appear if present. 
    The 1-$\sigma$ spectrum uncertainties
    are shown in gray, and the two red vertical dashed lines show the expected wavelength region of each emission band.
    Upper limits to H$_2$O and CO production rates were calculated using integrated flux measurements between the dotted red vertical lines (see Section \ref{sec:gas-production}).
    }
    \label{fig:h2o-co-spec}
\end{figure}

We used the extracted spectrum (Figure \ref{fig:jwst-spectrum}) to assess emission from H$_2$O, CO$_2$, and CO after contributions from dust and the nucleus were removed by fitting a third-order polynomial to the spectrum around each molecular emission band using a least-squares approach (see Figures \ref{fig:co2-spec} and \ref{fig:h2o-co-spec}) as described in Harrington \cite{harrington-pinto-2023}.

\section{Discussion}
\label{sec:discuss}

\subsection{Nucleus Radius and Color Estimates}
\label{sec:nuc_size}

We used the UTC 2022-08-21 and 2022-08-23 Gemini-N photometry data in Table \ref{tab:obs} to calculate 450P's effective spherical nucleus radius using the methods described in \cite{knight-2024-comets-III}.
We assumed a typical JFC nucleus geometric albedo in the $r'$ region of $p$ = 0.04 and a linear phase coefficient of $\beta$ = 0.047 mag/\degr \citep{knight-2024-comets-III} guided by the small nucleus size estimated during the previous apparition and the comet-like dust activity documented from 2003 to 2007  \citep{mazzotta-epifani-2006, mazzotta-epifani-2011, fernandez_2013}. 
We derived the following spherical nucleus radii for the two dates: $R_N$ = 2.1$\pm$0.1~km (2022-08-21) and $R_N$ = 2.3$\pm$0.1~km (2022-08-23).
The two radius estimates are close, but they do not agree exactly, likely because the nucleus was observed at different rotational phases. 
If the nucleus is nonspherical, its projected effective cross-section will change as it rotates, leading to slightly different radii derived from each set of data.

Both $p$ and $\beta$ are unknown for 450P, however, the range of known $\beta$ values is small, where \cite{knight-2024-comets-III} indicate 0.025 $\le \beta <$ 0.07 mag/$^{\circ}$ based on measured values for 24 comets, so the assumed value  of 0.047 is reasonable. 
At 450P's relatively large heliocentric distance of $\sim$ 7 au it is possible that fall-back dust grains may contain a non-negligible water ice content (see Section \ref{sec:coma-modeling}) and that this would increase the nucleus's albedo and consequently reduce the estimated radius.
For example, a geometric albedo of $p$ = 0.112 from \cite{romanishin-2018} yields $R_N$ = 1.3$\pm$0.1~km and $R_N$ = 1.4$\pm$0.1~km for the dates of 2022-08-21 and 2022-08-23, respectively.
Consequently, we estimate that the true nucleus radius is between the two limiting values of 1.3 to 2.3 km, derived by assuming geometric albedos that span the known Centaur range.
We therefore recommend using the average of these values, $R_N$ = 1.8$\pm$0.5~km.

Our radius is smaller than the {\it Spitzer}-derived estimate of $R_N \sim$ 3.5 km \citep{fernandez_2013}. 
450P had a prominent coma and tail during the Spitzer observations \citep{kelley_2013}, and their larger  nucleus size can be explained by residual dust coma likely being present in their analysis.
A non-spherical nucleus could also contribute to the differences. Regardless, it is clear that 450P possesses a small nucleus when compared to the known Centaur population \citep{stansberry_2008, bauer_2013, lellouch_2013}. 
However, we note that studies of the Centaur size-frequency distribution are likely biased to the larger members of the population \citep{fernandez2025}.

We measured the nucleus's surface color using photometry from the Gemini $g'$ and $i'$ images on UTC 2022-09-27 when the Centaur appeared to be inactive (Figure \ref{fig:stellar-profile-comparison}).
We found $g'-i'$ = 1.15$\pm$0.09~mag, which is one of the first bare nuclei color measurements of a previously active Centaur.
There are limited data in the literature presenting Centaur colors in the SDSS filter system for a direct comparison with 450P.
There are three Centaurs for which $g'-i'$ colors are reported: P/2019 LD$_2$ (ATLAS), $g'-i'$ = 0.77$\pm$0.06 \citep{bolin-LD2-2021AJ}; 2020 MK$_4$, $g'-i'$ = 0.59$\pm$0.04 mag \citep{de-la-fuente-2021A&A...649A..85D}; 174P/Echeclus, $g'-i'$ = 0.91$\pm$0.04~mag \citep{beck-2025}. 
450P's surface color is the reddest of the four, however, the color measurements of P/2019 LD$_2$ and 2020 MK$_4$ were obtained while a coma was present, so a direct comparison with the color of 450P is not straightforward.
Previous measurements of active Centaur nuclei colors (e.g., \cite{jewitt_2009, wong_2019, mazzotta-2018A&A...620A..93M}) used the Johnson-Cousins filter system and largely relied on observations obtained post-perihelion as objects recede toward their next aphelion, when low-level coma activity is more likely to persist.

One longstanding question is why active Centaurs appear preferentially within the gray/neutral population, which exhibits a bimodality including a red component \citep{peixinho-2025cent.book....5P}.
Proposed explanations include activity driven blanketing or thermally-driven material alterations that may transform red surfaces into gray/neutral ones \citep{jewitt_2009, wong_2019}.
450P's surface color is in this context interesting because it has recently migrated inward to its current cis-Saturnian orbit following its recent \ajump{}, where volatile-driven activity and thermally-driven surface processing increase.
The relatively red surface color measured for 450P, despite activity \citep{hahn_2006, mazzotta-epifani-2006, mazzotta-epifani-2011, kelley_2013}, may indicate that this surface-neutralization process is still at an early stage. 
If so, 450P provides an opportunity to observe the evolution of a Centaur surface in response to a hotter thermal environment following a recent inward migration. 
Should activity cease, follow-up observations to measure the bare nucleus color would provide a test of whether continued surface processing drives its nucleus toward a more gray/neutral color.

\subsection{Bulk Composition of Coma Dust}
\label{sec:coma-modeling}

To assess the bulk composition of 450P's coma we applied two kinds of spectral models to the JWST data: 
one which assumed the grains were large enough for a \citet{hapke2012} style model (previously applied to active Centaur comae in \citealt{kareta-2021}) and one in which the typical grain size was similar to or smaller than the wavelengths in question (e.g., utilizing Mie theory, previously applied to comet comae by \cite{yang2014}, \cite{protopapa-2021PSJ.....2..176P}, \cite{kareta_noonan_2023}, and others). 
For both models, we assumed that the two components that dominated the scattering cross section were amorphous carbon \citep{randm1991} and water ice \citep{mastrapa2008, mastrapa2009}.

For the Hapke models, we tested scenarios where the ice and dust grains are physically separated (``linear mixing") and ones where ice and dust are combined into single grains (``intimate mixing"). 
The models are fit to the data first by a simple $\chi^2$ minimization followed by a Markov Chain Monte Carlo (MCMC) approach (implemented through the \textit{emcee} package, \citealt{emcee2013}) using the least-squares optimized result as an initial guess. 
With all models, we assume that all light is coming from the solid particles in the coma with no nuclear contribution, which we return to later. 
A discussion of best practices and limitations in applying these kinds of models is available in \citet{firgard-2025PSJ.....6..184F}.

\begin{figure}[h!]
    \centering
    \includegraphics[width=0.75\textwidth]{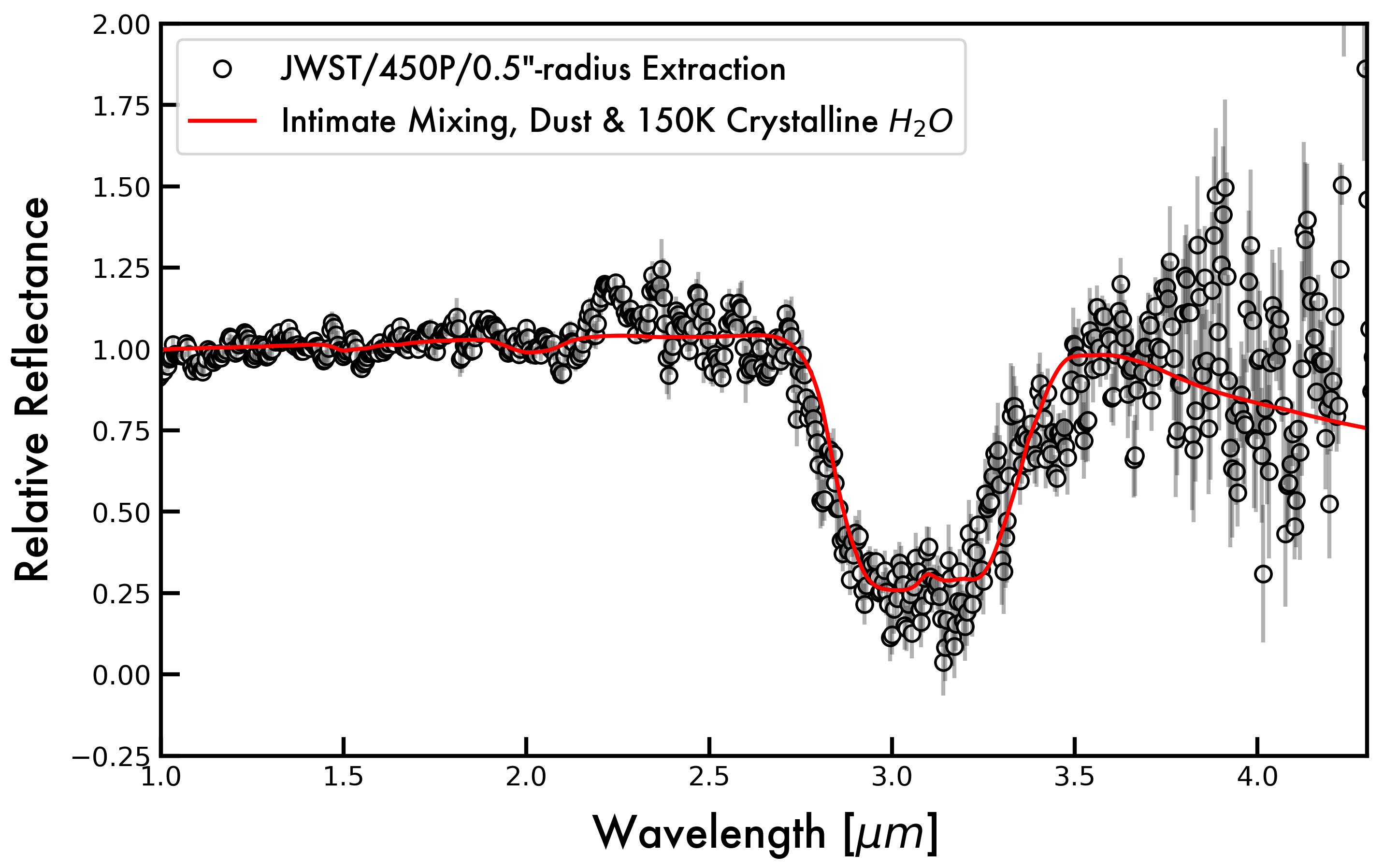}
    
    \caption{The reflectance spectrum of 450P (black unfilled circles with gray 1-$\sigma$ error bars) is compared with the maximum likelihood spectral model (red curve) which assumes intimately-mixed grains of amorphous carbon and crystalline water ice.
    This model is not very sensitive to ice temperatures, so while it assumes $T_{ice}=150$ K, ice temperatures within several tens of Kelvin fit the data similarly well.
    }
    \label{fig:spectral_model}
\end{figure}

Our modeling indicates that relatively large, on the order of microns, intimately-mixed grains are required to fit the depth and shape of the 3 $\mu$m absorption band of water. 
Warm ($>$ $\sim$ 100 K) crystalline ice provides a better fit than amorphous water ice or cooler samples of crystalline ice. 
We modeled the wavelength range of 1.1 $\mu$m $< \lambda <$ 4.2 $\mu$m where optical constants for water ice were measured and to exclude the strong CO$_2$ emission feature at $\sim$ 4.26 $\mu$m.  
We constrained the typical grain size in the inner coma to be $D_{eff.} = 5.9_{-0.5}^{+0.7}$ $\mu$m with a volumetric ice fraction of $f_{ice} = 33_{-1}^{+1}\%$ assuming 150 K crystalline ice. 
The upper- and lower-limits to $D_{eff.}$ and $f_{ice}$ correspond to the 16th and 84th percentiles of the distribution of parameters, and thus are broadly similar to $1\sigma$ error bars. 
Smaller grain sizes slightly correlated with higher ice fractions. 
The maximum likelihood model, with \deff{} and \fice{}, is shown in Figure \ref{fig:spectral_model}.

The results change by less than one standard deviation should slightly cooler ice temperatures be chosen (e.g., $\sim$ 120 - 140 K). 
Considering that the observations of 450P were taken at $R_H=7.16$ au, a slightly cooler ice temperature would be expected based on a simple blackbody temperature model. 
As can be seen from Figure \ref{fig:spectral_model}, the edges of the 3~$\mu$m band are not perfectly captured by this two component model. 
This, when combined with the ice temperature issue, suggests that the amorphous carbon is an incomplete descriptor of the underlying continuum. 
The ideal opaque to be used as an analog for the dust grains might have structure in the 3 $\mu$m region (the fit is poorest near $\sim$ 3.3--3.45~$\mu$m), but the SNR of the dataset is not high enough to discriminate between  alternative opaques.
Hydrated materials often have absorption features in this region, so if the dust was even partially hydrated, this might explain some of the ($\sim1\sigma$) challenges in modeling this wavelength range.

Another explanation for the under-prediction of reflectivity near 3.3--3.45 $\mu$m would be if 450P had weak (e.g., non-significant in the current dataset) emission or absorption from organic molecules in this wavelength range.
Considering the 3 $\mu$m absorption feature's depth, the spectral flux density in the region may be too low to help resolve the shape of the bottom of the band.
That said, the ``bump" in the spectrum at $\sim$ 3.1 $\mu$m is placed where the Fresnel peak from reflection from crystalline water ice is expected. 
If the overall uncertainty in the spectrum is dominated by photometric calibration rather than point-to-point scatter from on-target SNR, this indicates that the 3.1 $\mu$m bump is a real feature and provides further evidence for the substantial presence of crystalline water ice in relatively large ($\sim$ 6~$\mu$m) grains in 450P’s coma.

\subsection{Dust Production Rates} \label{sec:dust-production}

The brightness of dust in the coma at optical wavelengths can be used to calculate a mass loss rate.
Dust production rates can be estimated from the JWST data; however, such calculations are more complex and currently lack a standard methodology.
We estimated and removed the nucleus contributions from the 2023-07-30 and 2024-01-31 Gemini-N aperture photometry using the radius estimates derived in Section \ref{sec:nuc_size}.
The nucleus radius enters the analysis only through the adopted radius–albedo combination. 
The preferred value of $R_N =$ 1.8$\pm$ 0.5 km is an average of estimates obtained under different assumed geometric albedos and therefore does not correspond to a unique albedo. 
Equivalent nuclear fluxes are obtained using either $(R_N$ = 2.2~km, $p$ = 0.04) or $(R_N$ = 1.35~km, $p$ = 0.112), resulting in identical nucleus subtraction. 
The nucleus accounted for $\sim0.1$~mag of the total brightness, yielding coma apparent magnitudes of $m_c$ = 22.76$\pm$0.05 and $m_c$ = 21.60$\pm$0.04, respectively.  
The effects of uncertainties in our nucleus model parameters are likely to be small compared to the uncertainties in the assumptions needed to convert dust brightness to mass loss rates that follow.

We first quantify the dust coma using the $Af\rho$ parameter, which is the product of albedo, aperture filling factor, and aperture radius, and is proportional to the dust mass loss rate for comae in free expansion \citet{ahearn_1984,fink_2012}.  
It has units of length and can be used to compare comae to infer variations with time or dust properties.  
We corrected for phase angle using the Schleicher-Marcus phase function\footnote{\url{https://asteroid.lowell.edu/comet/dustphase/}} and derive $A(0\degr)f\rho$~=~48.0 and 94.0~cm for 2023-7-30 and 2024-01-31, respectively.

We converted $A(0\degr)f\rho$ to mass-loss rates using:

\begin{equation} \label{eq:dust_m_dot}
    \dot{M}_{\mathrm{dust}} = \frac{8 a \rho_d v_d (A(0\degr)f\rho)}{3 A_p},
\end{equation}

\noindent where $a$ is the effective radius of the assumed-spherical dust grains, $\rho_d$ is dust grain density, $v_d$ is dust grain radial expansion velocity projected onto the sky, and $A_p$ is geometric albedo of the dust grains.  
Uncertainties in effective dust size and expansion speed have the largest impacts on the mass loss rate.  
In principle, dust grain sizes may be sub-micrometer to centimeter, and the effective grain size depends on the size distribution.
Expansion speeds may also vary greatly, but tens of meters/sec to hundreds of meters/sec are commonly assumed.  
Here, we use our near-IR spectral modeling results (Section \ref{sec:coma-modeling}) as a nominal size ($a=3$~\micron) and assume  $\rho_d=1$~g~cm$^{-3}$, $v_d=50$~m~s$^{-1}$, and $A_p=0.05$.  
The resulting mass loss rates are 4 and 8~kg~s$^{-1}$ for for 2023-7-30 and 2024-01-31, respectively.

\subsection{Gas Production Rates: \cotwo{}, \water{}, and CO} \label{sec:gas-production}

While gas comae characterizations for comets have been undertaken for  decades \citep{bockelee-2004come.book, harrington_2022, biver-2024come.book..459B}, measurements for Centaurs have been limited until JWST \citep{fernandezvalenzuela2025}, with the exception of earlier millimeter-wavelength spectra of CO in 29P/Schwassmann-Wachmann~1, 95P/Chiron, and 174P/Echeclus \citep{senay-1994Natur.371..229S, womack-1999SoSyR..33..187W, kacper_2017} and \water{} and HCN in 29P \citep{bockelee-morvan-29P-2022A} and infrared detections of CO and upper limits of CO$_2$ in 29P \citep{ootsubo_2012_CO+CO2}.
Prior to this work, detections of \cotwo{} have been reported for only three Centaurs using JWST NIRSpec: 39P/Oterma \citep{harrington-pinto-2023}, 29P/Schwassmann-Wachmann 1 \citep{faggi-2024NatAs}, and 95P/Chiron \citep{pinilla-2024A&A-chiron}.

The shape of the CO$_2$ emission band (Figure \ref{fig:co2-spec}) constrains the molecule's rotational level population models.  
To retrieve the estimated gas production rate and its rotational temperature, we fit the continuum-subtracted spectrum with the Cometary Emission Model of the Planetary Spectrum Generator \citep{villanueva-2011-psg, villanueva-2018-psg}, using an outflow velocity, $\nu$, of 217 m s$^{-1}$ derived from equation $\nu$ = 580 $\times R_H^{-0.5}$ m/s from \cite{delsemme-1982come.coll...85D}, where $R_H$ is in units of au.
The production rate and rotational temperature were allowed to simultaneously vary, and the optimal fit was derived with least-squares fitting using \texttt{scipy}'s implementation of the Broyden, Fletcher, Goldfarb, and Shanno (BFGS) method.  
Model uncertainties were derived with an MCMC approach.  
The posterior probability distributions of the fitted parameters were nearly symmetric, so for simplicity we report the Bayesian uncertainties using effective 1-$\sigma$ errors based on the inner (68\%) confidence limits of the posterior distributions.   
The results are $Q_{\mathrm{CO_2}}$ = \cotwoproduction{}~molec.~s$^{-1}$, $T_\mathrm{rot}$ = 60$\pm$1~K, and the model spectrum is shown in Figure~\ref{fig:co2-spec}.
Comparable \cotwo{} rotational temperatures of $\sim$ 60~K have been reported for comet C/2024 E1 (Wierzchos; \cite{snodgrass-E1-2025MNRAS.541L...8S}) at $R_H$ = 7.2~au and for comet C/2017 K2 (PANSTARRS; \cite{woodward-K2-2025PSJ.....6..139W}) at 2.35~au.

For \water{} and CO upper limits, we estimated the integrated flux density of a simulated 3-$\sigma$ source following the methods described in \cite{ootsubo_2012_CO+CO2}, and the values are in Table \ref{tab:gas-rates}.
An outflow velocity of 217 m s$^{-1}$ was also used for these molecules.

\begin{deluxetable*}{cccccc}[h!]  \label{tab:gas-rates}
\tablecaption{Integrated band flux densities, column densities and production rates for 450P from 2023-09-03 JWST NIRSpec}
\tablecolumns{6}
\tablewidth{0pt}
\tablehead{
\colhead{Molecule} &
\colhead{$\lambda_x$} & 
\colhead{Integrated Flux} &
\colhead{g} &
\colhead{$ \langle N_x \rangle $} & 
\colhead{Q$_x$} \\
\colhead{} &
\colhead{$\mu$m} &
\colhead{W m$^{-2}$} &
\colhead{10$^{-4}$ s$^{-1}$} &
\colhead{molec. m$^{-2}$} & 
\colhead{molec. s$^{-1}$}
}

\startdata
H$_2$O ($\nu_3$) & 2.66 & $\le$ 6.1$\times$10$^{-22}$ & 2.85 &  $\le$ 1.2$\times10^{15}$ & $\le$ \htwooproduction{}  \\
CO$_2$ ($\nu_3$) & 4.26 & (2.14$\pm$0.02)$\times10^{-20}$ & 28.6 & (7.04$\pm$0.07)$\times10^{15}$ &  \cotwoproduction{}  \\
CO v(1–0) & 4.67 & $\leq$ 1.3$\times$10$^{-21}$ & 2.46 & $\leq$ 5.2$\times$10$^{15}$ & $\leq$ \coproduction{} \\
\enddata

\end{deluxetable*}

We calculated an upper-limit for the abundance ratio  $Q_{\mathrm{CO}}$/$Q_{\mathrm{CO}_2}$ $\le$ 0.7 and a lower-limit for $Q_{\mathrm{CO_2}}$/$Q_{\mathrm{H_2O} } \ge$ 5.8. To date, these production rate ratios were measured in only two other Centaurs: 29P/Schwassmann-Wachmann and 39P/Oterma (e.g., Figures~8 and~9 of \citealt{harrington-pinto-2023}). These  ratios are much closer to the values calculated for 39P than for 29P. Comparing orbits of these bodies, 39P had a relatively recent history of a closer orbit consistent with that of a JFC before its current orbit as a Centaur, while 29P  probably never has been on a JFC orbit. It is interesting that 450P, which has not to our knowledge had a close-in orbit to the Sun, more closely resembles 39P than 29P for the CO and CO$_2$ production rate ratios. 
Of course, these coma abundances may arise from different formation environments. 
Unfortunately, the dataset of Centaurs is too small to warrant further interpretations.

\subsection{Thermal Modeling: Probing the Interior of 450P's Nucleus Over Time}
\label{sec:therm}

\begin{figure}[ht]
\centering
\includegraphics[width=0.58\textwidth]{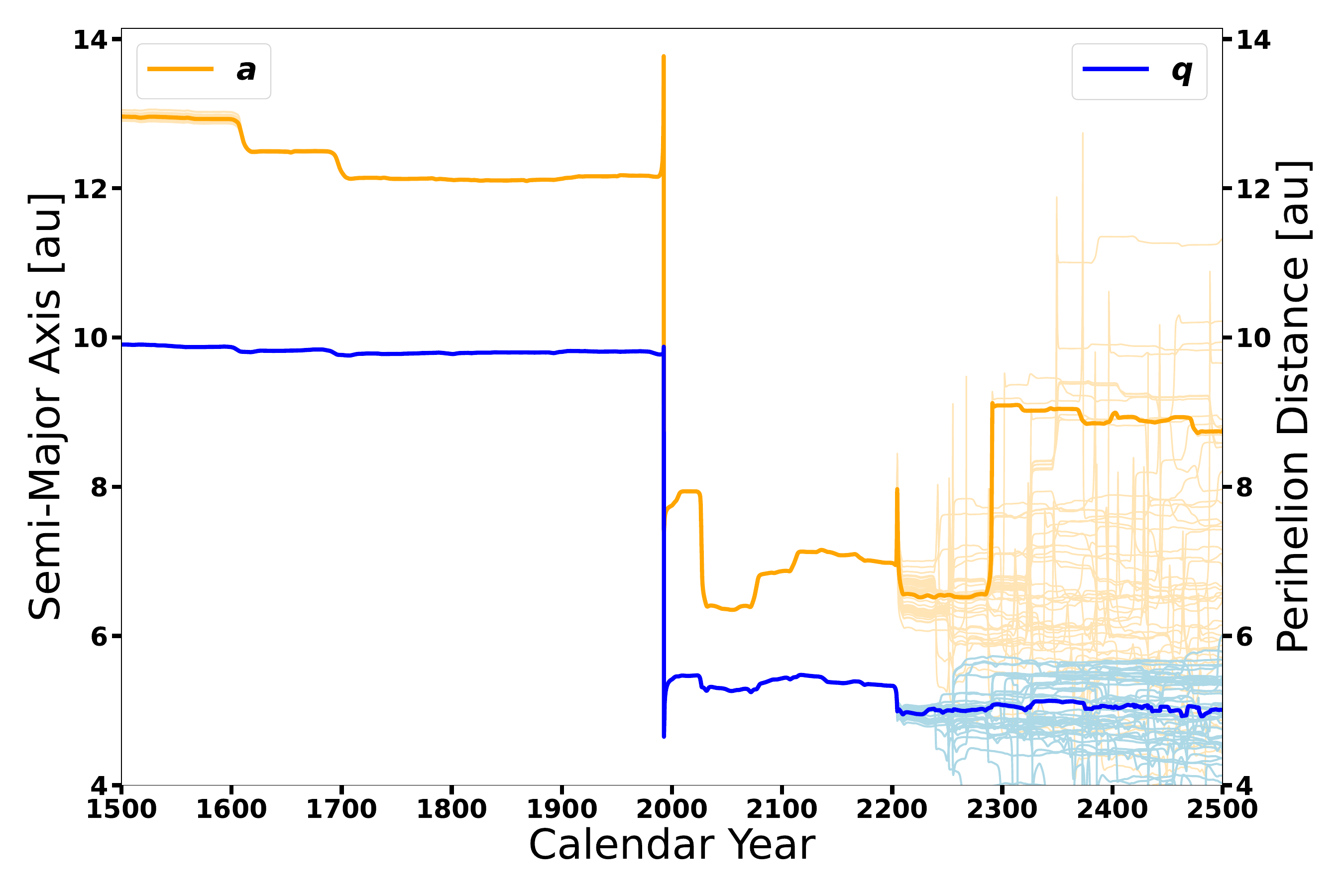}
\caption{The evolution of 450P's perihelion distance ($q$, blue curve) and semi-major axis ($a$, gold/orange curve) from 1500 - 2500 AD using 50 clone orbital integrations.
The recent \ajump\ after the 1992 close approach to Saturn is easily seen by the rapid decrease in both $a$ and $q$.}
\label{fig:orbit-history-clones}
\end{figure}

Here, we explore how our observations may constrain thermal models of 450P's interior, including plausible activity mechanisms. 
We implemented a simple thermal model following the approach presented in \cite{lilly-2024ApJ} to compute the solar heating received by 450P's nucleus over the past and future 500 years.
The orbital inputs were determined through a series of backwards numerical orbit integrations for 50 clones using the software REBOUND \citep{rein-2012}.
The simulations for the orbit's semi-major axis ($a$) and perihelion distance ($q$) are shown in Figure~\ref{fig:orbit-history-clones} for the 1,000-year time interval (1500~CE--2500~CE) and indicate that the orbital history of 450P is well constrained since $\sim$ 1500 CE.
Compared to earlier analyses (e.g., \citealt{hahn_2006}), Figure~\ref{fig:orbit-history-clones} extends the temporal baseline both backward and forward in time and explicitly includes the evolution of the semi-major axis, providing a more complete view of the object’s long-term dynamical behavior.
The most recent $a$-jump is easily identified by the decrease in both $a$ and $q$ in 1992 CE, as shown in Figure \ref{fig:orbit-history-clones} and  previously identified in \cite{hahn_2006} and \cite{lilly-2024ApJ}.

This model reveals how the orbital change altered 450P's surface and subsurface thermal state. 
Figure \ref{fig:therm} presents the modeled evolution from 1870~CE--2030~CE, including the sub-solar surface temperature and heating to a depth of $\sim$ 200~m.
Before the most recent \ajump{}, surface temperatures remained relatively stable at $\sim$ 105-125~K; afterward, the reduced heliocentric distance led to a pronounced increase in solar heating, with surface temperatures rising to roughly 125-170~K.
The interior follows the same overall trend with a delayed response expected from thermal diffusion.
While the thermal model is simpler than the ones \citep{capri-2009}, the ability to include a \cotwo{} gas detection provides an advantage by providing more modeling constraints.

Our approach for including the orbital history for 450P ignores the possibility that it possessed an orbit with excursions interior to 15.9~au prior to 1500~CE, which would have heated the nucleus and its interior to temperatures higher than what our model assumes.
On the other hand, the absence of nucleus rotation in the model may lead to an overestimation of localized heating at the subsolar point, since it receives continuous insolation without diurnal cycling. 
Thus, our model may be viewed as a lower limit of the heating received because it does not include past epochs when 450P was interior to Saturn.

\subsubsection{Constraints on 450P Activity from \cotwo{} Detection}

The detection of \cotwo{} gas provides a strong observational constraint on the activity mechanism at 450P.
While non-detections of other volatile species provide useful consistency checks, the \cotwo{} detection directly links the observed activity to thermal processes occurring within the nucleus.

\begin{figure}[ht]
    \centering
    \includegraphics[width=0.5\textwidth]{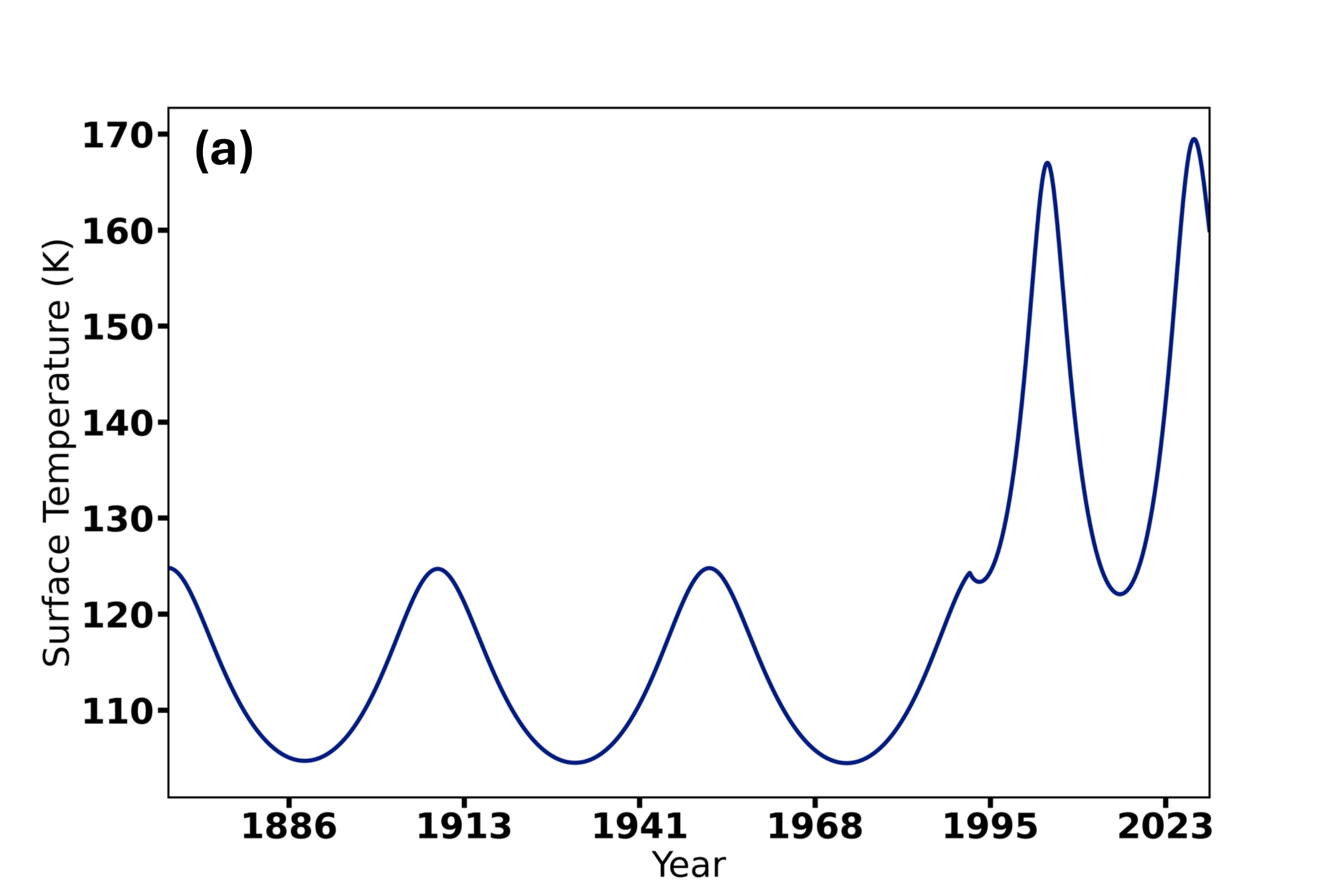}~
    \includegraphics[width=0.5\textwidth]{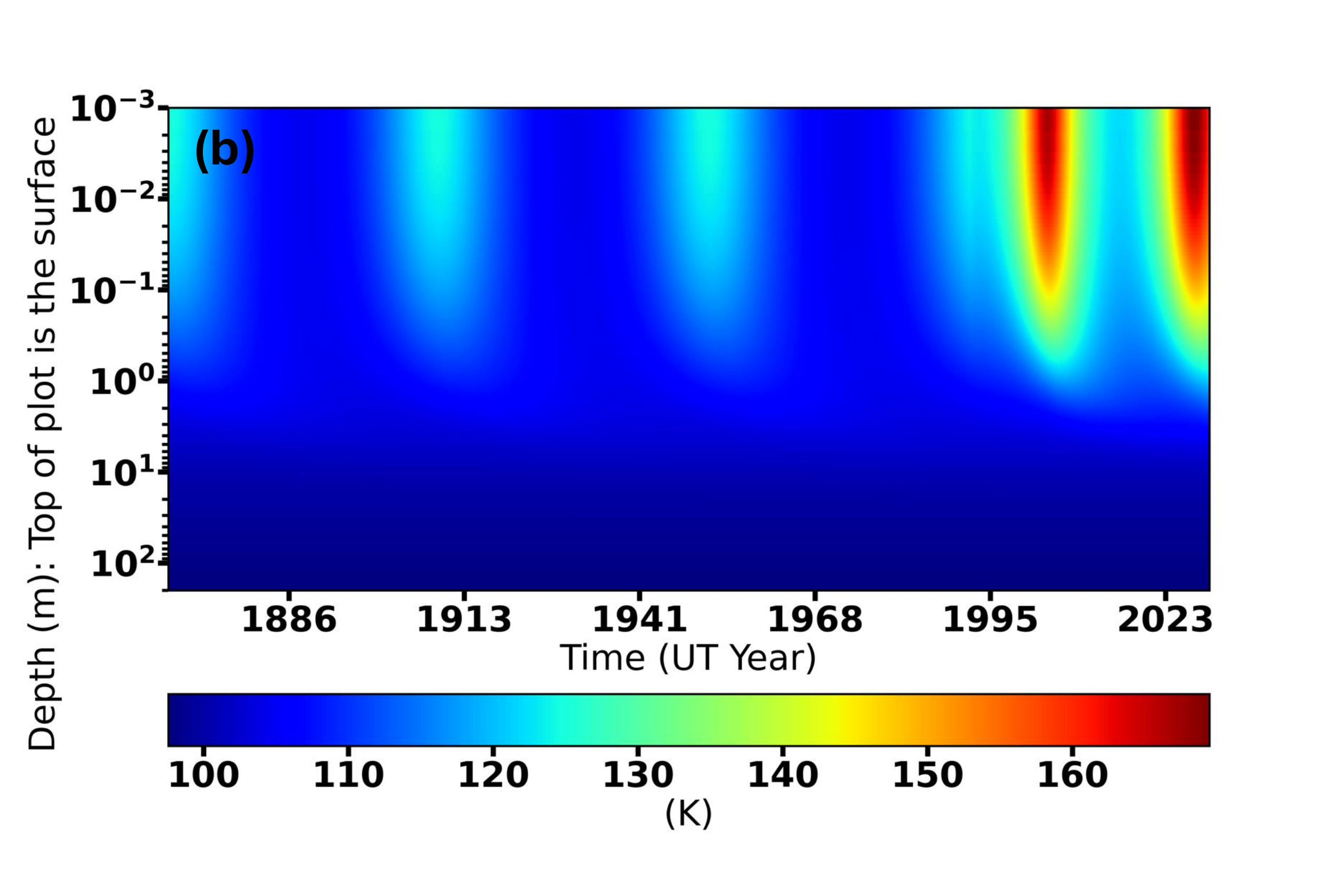}
    \includegraphics[width=0.5\textwidth]{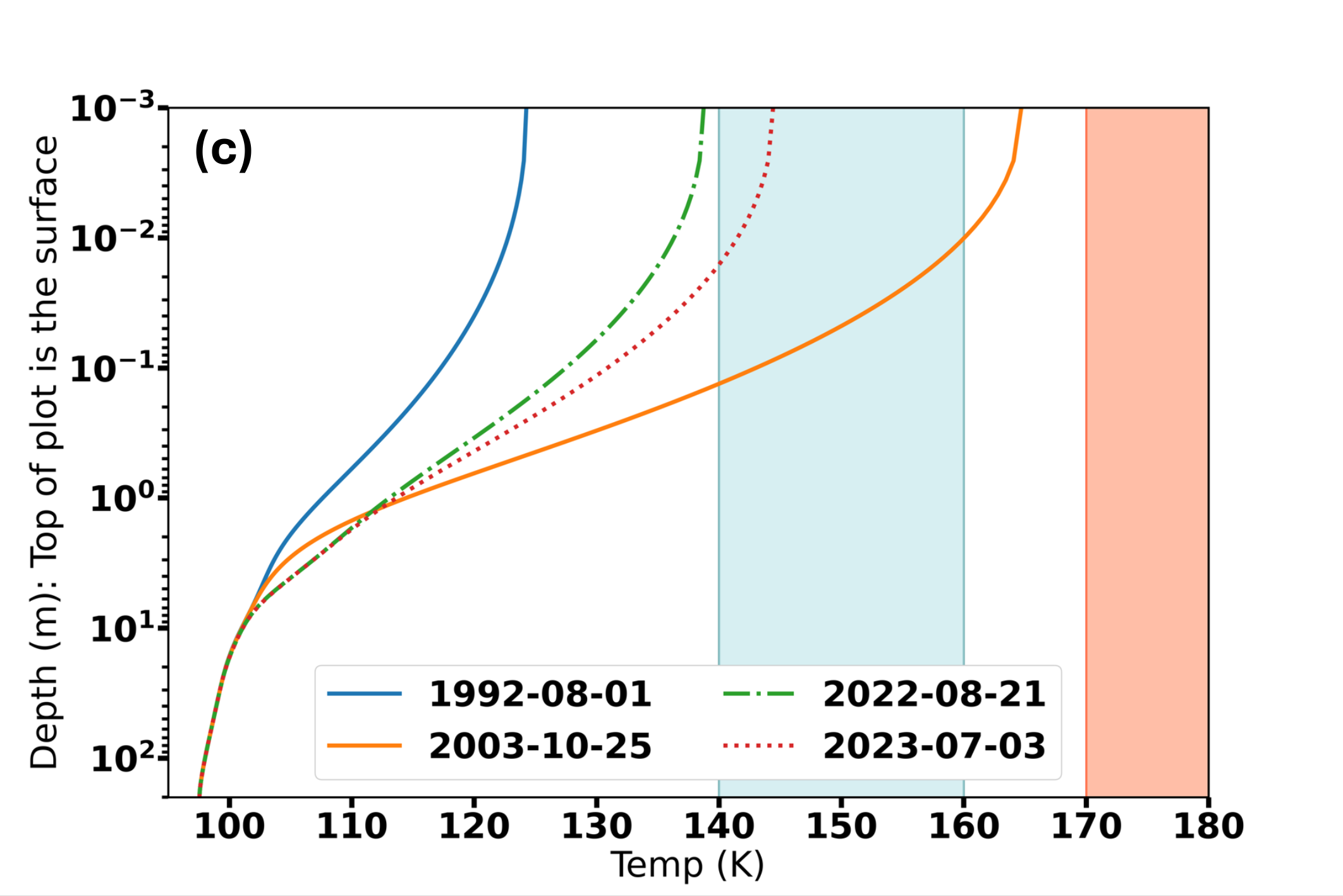}
    
    \caption{These panels display the time evolution of the (a) sub-solar surface temperature, (b) radial profile of nucleus's interior temperature, and (c) selected nucleus interior radial temperature profiles for dates when dust production estimates are available.
    The thermal effects due to the \ajump{} in 1992 are easily seen by the increased temperatures that occurred shortly afterward.
    In Panel (c), the cyan shaded region between temperatures [140 K, 160 K] represents temperatures where crystallization of AWI water ice occurs efficiently and with it the possibility for the release of trapped \cotwo{}.
    The light-red shaded region displays temperatures where the sublimation of crystalline water ice begins to dominate.}
    \label{fig:therm}
\end{figure}

Figure~\ref{fig:therm}, panel~(c), shows radial profiles of temperature at orbital positions with heliocentric distances corresponding to: 
shortly before the 1992 close Saturn encounter; 
the 2003-04-03 discovery when activity was first detected (Figure \ref{fig:orbit}, marker 2); 
the 2022-08-21 recovery, point-source detection where no activity was detected (Figure \ref{fig:orbit}, marker 10); 
and the 2023-07-03 pre-JWST observations (Figure \ref{fig:orbit}, marker 11).
The radial profiles of these positions were chosen to help inform plausible depths in the nucleus interior where temperatures are compatible with the onset of 450P's detected dust and \cotwo{} gas emission.
The two scenarios we consider for the release of \cotwo{} include: (1) the sublimation of pure \cotwo{} ice and (2) the release of trapped \cotwo{} from cavities in a porous amorphous water ice (AWI).

Our model shows the interior of 450P reaches temperatures exceeding the sublimation activation temperature for \cotwo{} ($\sim$ 70-80 k) to depths $> 200$ m well before the 1992~CE \ajump.
Thus, it is likely that pure \cotwo{} ice deposits near 450P's cometographic equatorial regions would have been depleted by the time activity was first detected in the 2003-10-15 discovery data.
However, a pure \cotwo{} ice sublimation activity driver cannot be ruled out as a gas coma originating from high cometographic latitudes.  
Near-polar regions on 450P's nucleus could still retain sufficient \cotwo{} ice to drive the activity given a sufficiently low nucleus obliquity.
The \cotwo{} morphology in the coma later seen in the JWST data (see Figure \ref{fig:ccoma-morpho}), however, is inconsistent with a source region at a high cometographic latitude.
A source at such a high latitude would produce a more asymmetric, fan-shaped morphology, providing more evidence for another mechanism of \cotwo{}'s release than sublimation of a pure ice source.

Alternatively, our model indicates that internal temperatures were not high enough for vigorous crystallization before the 1992~CE \ajump{}.
While AWI undergoes crystallization at temperatures below $\sim$140~K, laboratory experiments show that the release efficiency of trapped volatiles is reduced between $\sim$~80-140~K then increases again near $\sim$~140~K, when gas emission becomes more efficient during the AWI hexagonal phase transition \citep{prialnik-2024-comets-III}.
The radial profile for 2003-10-25 shows that 160 K is achieved at the surface and the interior is heated to $>$ 140 K down to $\sim$~13~cm.
Thus, activity driven by the release of trapped gases from AWI is a plausible explanation for the detected dust activity in the 2003-10-25 data; however, no gas species were detected at the time, so we cannot determine whether \cotwo{} was the driving volatile.

The radial profile from the 2022-08-21 Gemini observations, while detected as a point source, indicates that 450P likely was too cold to be active through \cotwo{}'s emission during the AWI phase transition that occurs at $\sim$~140~K \citep{prialnik-2024-comets-III}.
However, the 2023-07-03 Gemini images and 2023-09-03 JWST data were acquired when the model indicates that 450P's interior had reached at least 140 K to a depth of several cm. 
Thus, AWI crystallization is a viable activity driving mechanism for the release of the observed \cotwo{} emission and subsequent dust lofting at 450P.

\subsubsection{Consistency of CO and H$_2$O Non-Detections with CO$_2$-Driven Activity}

Having established that \cotwo{} release is consistent with thermally driven activation through AWI crystallization after its 1992 CE \ajump{}, we now consider whether the absence of other volatiles detections is compatible with this interpretation.
In particular, CO and \water{} non-detections provide additional context for volatile stratification and phase evolution within the nucleus's interior.
The following discussion should therefore be interpreted primarily as tests of consistency with this AWI-crystallization, \cotwo-driven activity scenario rather than proposing independent activity driving mechanisms.

The lack of a detected CO gas coma may be inconsistent with scenarios in which the \cotwo{} originates from release during the crystallization of AWI.
One possibility for the lack of CO production compared to CO$_2$ is that the near-surface layers of AWI have already undergone a few phase changes where the bulk of the trapped CO, and potentially CH$_4$, were lost. 
At $\sim$ 60 K AWI changes from a high-density to a low-density phase, where the desorption of CO may preferentially occur due to its lower binding energy when compared to \cotwo{} \citep{prialnik-2024-comets-III}.
This argument has been proposed to explain CH$_4$ in the coma of the Centaur Chiron near its aphelion \citep{pinilla-2024A&A-chiron}.
So, while there may still be AWI present in the near surface layers of 450P, those layers are likely depleted in volatiles less stable than \cotwo{}, although they may be stored in the deeper interior. 
While some CO release cannot be ruled out, its production rate is likely too low to be detected in the available NIRSpec data.
Another possibility is that 450P could have formed in a region with a higher abundance of \cotwo{} during the AWI's growth on dust grains.

The lack of detected \water{} gas fluorescence in the coma of 450P at $R_H$ $\sim$ 7 au is not surprising as Figure \ref{fig:therm} indicates that the peak temperature reached by its nucleus near perihelion does not exceed 170 K. 
This is below the $\sim$ 180~K temperature when the onset of vigorous water ice sublimation begins \citep{meech_svoren_2004, prialnik-2024-comets-III}. 
Also, the $\sim$170~K temperature is reached only at the sub-solar point on 450P’s nucleus. 
Temperatures in the interior beneath this region (see Figure~\ref{fig:therm}) and across surface areas away from the sub-solar point decrease rapidly, such that the effective area of water ice exposed to higher temperatures is a small fraction of the nucleus’s total area.
So, a detectable amount of \water{} gas emission in the coma due to \water{} sublimation from the nucleus is unexpected.

Alternatively, dust grains in 450P's coma do appear to have a relatively high abundance of water ice content based on the spectral modeling included in Section \ref{sec:coma-modeling}, so an \water{} gas emission in the coma could be produced by the sublimation of icy grains from an extended coma source. 
Such an extended source of \water{} was seen with the Centaur 29P/Schwassmann-Wachmann~1 \citep{bockelee-morvan-29P-2022A}.
To assess this possibility at 450P, we calculate the total effective \water{} production from an extended coma source by estimating the sublimation rate of icy grains with an effective radius $a = 3 \mu$m, a volume water ice fraction of \fice{} at a range of plausible grain temperatures.
While these values are representative of the best-fit spectral modeling results, there exists a known degeneracy between grain size and ice fraction, such that larger grains with lower ice fractions could produce similar spectral properties. 
Such grain populations would produce comparable or lower total water production rates, and therefore would not materially alter the conclusions of our analysis.
The total number of dust grains in the coma contained within the photometric aperture used for the JWST spectrum's extraction was estimated using the $Af\rho$ measurement from the UTC 2023-07-30 Gemini data.
We adopt the Gemini measurement for this first-order estimate of the water production associated with an extended icy grain coma. 
While a dust production rate could in principle be derived from the JWST observations, such an analysis is non-trivial and would not materially alter the conclusions inferred from the more standard dust production estimate obtained from the Gemini broadband imaging data.
The total number of grains ($N_{\mathrm{dust}}$ = 6.25$\times 10^{17}$ grains) is given by

\begin{equation}
    N_{\mathrm{dust}} = \frac{ Af\rho \times \rho }{ A a^2 },  
\end{equation}

\noindent where the variables included are defined in Section \ref{sec:dust-production}.
For the calculation we used the methods from \cite{beer_2006}, where the derived water ice fraction \fice{} was used to calculate a dust grain mass fraction for water ice $X = 0.25$ (where $X$ = 1 for pure water ice), assuming a water ice density $\rho_{w}$ = 900 kg/m$^3$ and for dust content $\rho_d$ = 500 kg/m$^3$. 
The dust grain temperatures at $R_H$ = 7.16~au are likely colder than the best-fit value of 150 K derived from the coma modeling in Section \ref{sec:coma-modeling}.
As mentioned earlier, the arrived at 150 K temperature is likely due to the simplification of dust grain compositions used in the model being limited to two constituents: amorphous carbon and water ice.
Lower grain temperatures are further supported, where a total expected water production rate is on the order of $10^{28}$ molec. s$^{-1}$ using 150 K, well above the upper-limit rate of $\le$ \htwooproduction\ molec. s$^{-1}$.
Results from \cite{beer_2006}, Figures 3-5 suggest that grains of $\sim$ 10 - 20 $\mu$m and with $X$ = 0.25 should be lower than 140~K.
\cite{bockelee_2017} indicate that grains at $R_H = 7.16$~au should be at an equilibrium temperature of $\sim$~100~K (i.e., using $T_{equi.} = 278 \times R_H^{-0.5}$~K), however grains are somewhat hotter than this equilibrium temperature by $\sim$~20~\% due to inefficient radiation of thermal energy from micron-sized and smaller grains \citep{beer_2006, hanner_1997}.
Using slightly elevated grain temperatures of $\sim$120~K, we derive a water production rate of $Q_{\mathrm{H_2O}}$ on the order of $10^{24}$ molec. s$^{-1}$, consistent with the upper-limit water production rate estimated from the NIRSpec data.

\subsubsection{Implications for Volatile Retention in 450P and Recently Activated Centaurs}

Studies indicate that small km-sized Centaurs may be depleted in AWI throughout their nuclei due to the long-term, increased warming experienced over the typical $\sim$~10~Myr migration to an orbit interior to Saturn's \citep{lisse-2022PSJ-SW1-AWI, guilbert-2023,kokotanekova2025}. 
There are alternative scenarios though where an individual Centaur could have a rapid inward migration ($\sim$ 0.5 Myr) into a cis-Saturnian orbit, receiving less integrated heating, where AWI may be preserved \citep{gkotsinas-2022ApJ...928...43G, guilbert-2023}.
The combination of 450P's orbital thermal history, interior temperature evolution, and \cotwo{} morphology in the coma collectively favor activity driven by the AWI phase transition rather than sublimation of a pure \cotwo{} ice reservoir. 
Thus, the constraints on 450P's small nucleus size may have implications for the long-term survivability of AWI in its nucleus, and Centaurs in general, depending on the timescale of its recent inward migration.

The limited gas detection acquired to-date for 450P (one epoch from JWST) prevents confidence from being drawn from a higher level of thermophysical modeling \citep{prialnik-2004come.book..359P, huebner-2006} as it would likely be under-constrained.
Future long-term gas and dust production monitoring using JWST would provide more modeling constraints and enable a considerable improvement into understanding the interior of 450P.
We emphasize that 450P is a high priority target for such a study due to its recent \ajump{} and having experienced only one complete orbit with decreased $a$; it is now only beginning its second inbound perihelion passage with activity detected.
A long baseline of JWST observations would allow reliable measures of 450P's \cotwo{} and dust production rates while simultaneously monitoring for the onset of any detectable \water{} and/or CO gas emission in the coma.

\section{Conclusions} \label{sec:conclusions}

Our analysis of the relatively faint but active Centaur 450P with JWST and Gemini-N provides a comprehensive characterization of this object. 
450P had a close approach to Saturn in 1992, which changed its orbit and triggered measurable activity. It is heading toward another orbit-changing event with Jupiter in 2026 which we anticipate may lead to increased activity. 
We summarize key findings for 450P and our interpretations here:

\begin{itemize}

    \item After 15 years of being undetected, 450P was recovered using the Gemini-N Observatory on UTC 2022-08-21 and 2022-08-23 and appeared as a point source. These images are possibly the first images of 450's bare nucleus.
    Using assumed geometric albedos on the edges of the known Centaur population (0.04, 0.11) we derive an effective spherical radius of $R_N$~=~1.8$\pm$0.5~km. This marks 450P as one of the smallest Centaurs studied.

    \item Imaging data from Gemini-N on UTC 2022-09-27 were used to measure the nucleus color of g$'$~-~i$'$~=~1.15$\pm$0.09, which places 450P on the red end of the neutral/gray Centaur color group. 
    This color is also slightly redder than a typical comet nucleus and is consistent with a surface that has experienced comparatively less solar-driven thermal processing.

    \item Monitoring with Gemini-N as 450P approaches its UTC 2027-07-25 perihelion passage shows an onset of coma activity between 7.23~au~$\le~R_H~\le$~7.83~au. 
    During this time, the dust produced appeared to almost double from $Af\rho~=~36~\pm~2$~cm on UTC 2023-07-30 to $Af\rho~=~69~\pm~2$~cm on UTC 2024-01-31, which corresponds to dust production rates of $\sim$~4 to $\sim$~8~kg~s$^{-1}$.

    \item JWST NIRSpec IFU Prism mode spectra were acquired on UTC 2023-09-03 when 450P displayed a coma consisting of dust and \cotwo{} gas. 
    Emission from \cotwo{} was strongly detected. 
     No emission features from a \water{} or CO gas source were detected. 
    We calculate production rates of  $Q_{\mathrm{CO_2}}$ = \cotwoproduction{} molec. s$^{-1}$, $Q_{\mathrm{H_2O}} \le$ \htwooproduction{} molec. s$^{-1}$, and $Q_{\mathrm{CO}} \le$ \coproduction{} molec. s$^{-1}$.

    \item The \cotwo{} gas morphology in the coma appeared azimuthally symmetric. 
    Given the low solar phase angle during the observations, this morphology is consistent with emission from the nucleus near the subsolar point.
    In contrast, the distribution of dust grains in the coma appeared elongated, with a tailward extension consistent with solar radiation pressure acting on the dust grains.

    \item The JWST reflectance spectrum appears to have absorption features at 2.0 and 3.0 $\mu$m which we attribute to water ice. The spectrum is best fit by a Hapke-style model with an intimate mixture dominated by large \deff{} dust grains with a volumetric ice fraction of \fice{}.
    A subtle feature at 3.1 $\mu$m is at the location of the crystalline \water{} Fresnel peak and provides evidence that crystalline water ice is also present in larger grains in 450P's coma.

    \item A simple thermal model that incorporated 450P's well-constrained orbital history from $\sim$~1500~CE in combination with its observed activity behavior indicates that is activity is plausibly explained by the release of \cotwo{} from an amorphous water-ice undergoing crystallization between 140~K and 160~K.

\end{itemize}

%% IMPORTANT! The old "\acknowledgment" command has be depreciated. It was
%% not robust enough to handle our new dual anonymous review requirements and
%% thus been replaced with the acknowledgment environment. If you try to 
%% compile with \acknowledgment you will get an error print to the screen
%% and in the compiled pdf.
%% 
%% Also note that the acknowledgment environment does not support long amounts of text. If you have a lot of people and institutions to acknowledge, do not use this command. Instead, create a new \section{Acknowledgments}.

%\begin{acknowledgments}

\section*{Acknowledgments}

This work is based in part on observations made with the NASA/ESA/CSA James Webb Space Telescope. 
The data were obtained from the Mikulski Archive for Space Telescopes at the Space Telescope Science Institute, which is operated by the Association of Universities for Research in Astronomy, Inc., under NASA contract NAS 5-03127 for JWST.
These observations are associated with program \#2416 and can be accessed via \dataset[doi: 10.17909/2vtm-st63]{https://doi.org/10.17909/2vtm-st63}. 
Support for program \#2416 was provided by NASA through a grant from the Space Telescope Science Institute, which is operated by the Association of Universities for Research in Astronomy, Inc., under NASA contract NAS 5-03127.

This work is based in part on observations obtained at the international Gemini Observatory (Program IDs: GN-2021A-FT-214, GN-2022A-FT-111, GN-2022B-LP-203), a program of NSF’s NOIRLab, which is managed by the Association of Universities for Research in Astronomy (AURA) under a cooperative agreement with the National Science Foundation on behalf of the Gemini Observatory partnership: the National Science Foundation (United States), National Research Council (Canada), Agencia Nacional de Investigaci\'{o}n y Desarrollo (Chile), Ministerio de Ciencia, Tecnolog\'{i}a e Innovaci\'{o}n (Argentina), Minist\'{e}rio da Ci\^{e}ncia, Tecnologia, Inova\c{c}\~{o}es e Comunica\c{c}\~{o}es (Brazil), and Korea Astronomy and Space Science Institute (Republic of Korea).

This work was enabled by observations made from the Gemini North telescope, located within the Maunakea Science Reserve and adjacent to the summit of Maunakea. We are grateful for the privilege of observing the Universe from a place that is unique in both its astronomical quality and its cultural significance.

The authors thank the director and staff of the international Gemini Observatory for their quick turnaround in approving and completing the follow-up imaging to confirm 450P's recovery. 
This project would not have been possible without their hard work. The authors would like to thank the two anonymous reviewers for their feedback and improving the quality of the manuscript.

C.A.S. additionally acknowledges funding from the NASA SSO Prog. Id: \#80NSSC23K0678 (PI: C. Schambeau) and the Florida Space Research Initiative for support of this work.

%\end{acknowledgments}

%% To help institutions obtain information on the effectiveness of their 
%% telescopes the AAS Journals has created a group of keywords for telescope 
%% facilities.
%
%% Following the acknowledgments section, use the following syntax and the
%% \facility{} or \facilities{} macros to list the keywords of facilities used 
%% in the research for the paper.  Each keyword is check against the master 
%% list during copy editing.  Individual instruments can be provided in 
%% parentheses, after the keyword, but they are not verified.

\vspace{5mm}
\facilities{Gemini North (GMOS-N), JWST (NIRSPEC)}

%% Similar to \facility{}, there is the optional \software command to allow 
%% authors a place to specify which programs were used during the creation of 
%% the manuscript. Authors should list each code and include either a
%% citation or url to the code inside ()s when available.

\software{\texttt{astropy} \citep{astropy_2013A},
          \texttt{astroquery} \citep{ginsburg-2019AJ-astroquery},
          \texttt{ccdproc} \citep{craig-2017},
          DRAGONS \citep{labrie-2019-dragons},
          \texttt{emcee} \citep{forman-2020-emcee},
          \texttt{photutils} \citep{larry_bradley_2024_13989456},
          REBOUND \citep{rein-2012},
          \texttt{scipy} \citep{virtanen20-scipy},
}

%% For this sample we use BibTeX plus aasjournals.bst to generate the
%% the bibliography. The sample631.bib file was populated from ADS. To
%% get the citations to show in the compiled file do the following:
%%
%% pdflatex sample631.tex
%% bibtext sample631
%% pdflatex sample631.tex
%% pdflatex sample631.tex

%\appendix

\bibliography{schambeau_refs}{}
\bibliographystyle{aasjournal}

%% This command is needed to show the entire author+affiliation list when
%% the collaboration and author truncation commands are used.  It has to
%% go at the end of the manuscript.
%\allauthors

%% Include this line if you are using the \added, \replaced, \deleted
%% commands to see a summary list of all changes at the end of the article.
%\listofchanges

\end{document}